\documentclass[apj]{emulateapj}
\usepackage{amssymb}
\usepackage{multirow}
\usepackage{graphicx}
\usepackage{subfigure}

\shorttitle{Star Formation \& Chemical Evolution in the Smallest Galaxies}
\shortauthors{Webster, Sutherland \& Bland-Hawthorn}

\newcommand{\comments}[1]{}

\begin{document}

\title{Ultrafaint Dwarfs - Star Formation and Chemical Evolution in the Smallest Galaxies}

\author{David Webster}
\affil{Sydney Institute for Astronomy, School of Physics, University of Sydney, NSW 2006, Australia}

\author{Ralph Sutherland}
\affil{Research School of Astronomy \& Astrophysics, Australian National University, Cotter Rd,
Weston, ACT 2611, Australia}

\author{Joss Bland-Hawthorn}
\affil{Sydney Institute for Astronomy, School of Physics, University of Sydney, NSW 2006, Australia}
\email{d.webster@physics.usyd.edu.au}

\begin{abstract}

In earlier work we showed that a dark matter halo with a virial mass of $10^7$~M$_\odot$ can retain a large percentage of its baryons in the face 
of the pre-ionization phase and supernova (SN) explosion from a $25~$M$_\odot$ star. Here we expand on the results of that work, investigating the star formation and chemical evolution 
of the system beyond the first SN. 
In a galaxy with a mass $M_{\rm{vir}} = 10^7~$M$_\odot$, sufficient gas is retained by the potential for a second period of star formation to occur. The impact of a central explosion is found to be much stronger than that of an off-center explosion both in blowing out the gas and in enriching it, as in the off-center case most of the SN energy and metals escape into the intergalactic medium. We model the star formation and metallicity given the assumption that stars form for 100, 200, 400 and 600 Myr and discuss the results in the context of recent observations of very low mass galaxies. We show that we can 
account for most features of the observed relationship between [$\alpha$/Fe] and [Fe/H] in ultra-faint dwarf galaxies with the assumption that the systems formed at a low mass, rather than being remnants of much larger systems.
\end{abstract}


\section{Introduction}

The origin of the smallest galaxies (M$_{300} \leq 10^7$~M$_\odot$) is an unresolved question. Some are likely to be remnants of much larger ($\sim 10^9$~M$_\odot$) systems 
that have been stripped of a large percentage of their baryons and dark matter \citep{mayer01}, but others may have formed as low-mass systems. The discovery 
of examples of this latter class of systems would boost our understanding of the early baryonic systems as they are likely to have preserved chemical
signatures of the first generations of stars. 
Small systems are simpler and are likely to have undergone fewer enrichment events, leaving these signatures
relatively intact \citep{bland10,karlsson12,frebel12,karlsson13}.

The Sloan Digital Sky Survey \citep{york00} discovered a number of very faint Local Group dwarf galaxies. These galaxies have luminosity $L_{\rm{tot}} 
< 10^5$~L$_\odot$ and are known as ultrafaint dwarfs (UFDs). The halo masses of these systems are uncertain and highly model-dependent, however the 
masses within the half-light radius are better constrained. \citet{wolf10} found that the half-light masses of the UFDs ranged from $6\times10^5$ to $1.2\times10^7$~M$_\odot$. Segue 2 \citep{belokurov09} contains less than $1.5\times 10^5$~M$_\odot$ within the half-light radius.

Early UFD studies such as \citet{simon07} studied chemistry using [Fe/H] histograms, but more recent works \citep{frebel12, vargas13, kirby13} have 
added an extra dimension
through observations of [$\alpha$/Fe]. \citet{tinsley79} suggested that enhanced [$\alpha$/Fe] in halo stars could be explained by the time delay between 
Type II and Type Ia supernovae (SNe) and this idea is now widely
used in chemical modeling. Type II SNe eject many more alpha elements relative to iron than Type Ia and can occur early in the lifetime of a galaxy because their
progenitors are short-lived high-mass stars, while Type Ia SNe are delayed until lower mass stars have evolved. This means that the enrichment is initially dominated
by Type II SNe such that $[\alpha$/Fe] is initially high, but declines after $\sim100~$Myr due to the effect of Type Ia SNe. The value of [Fe/H] at which 
[$\alpha$/Fe] begins to decline is linked to the number of Type II SNe in the first 100~Myr and is therefore an approximate measure of the star formation rate 
in a system. If there is no decline, it is likely that star formation lasted less than 100~Myr. However, as we will show, the stochastic nature of star formation
 in small systems means that the average time between Type Ia 
SNe can be as much as 50~Myr and it is therefore possible for a system to form stars for significantly
longer than 100~Myr without a Type Ia SN.

\citet{frebel12} investigated six UFDs and found that Ursa Major II was a good candidate for having no Type Ia enrichment, while Coma Berenices and Bootes I were
reasonable candidates given uncertainties in r-process predictions. \citet{vargas13} found that Segue I and Ursa Major II do not show any low [$\alpha$/Fe] stars
and, therefore, star formation likely lasted less than 100~Myr. However, five other UFDs do show a decline in [$\alpha$/Fe] with increasing [Fe/H]  and therefore 
likely had longer star formation histories. While total halo masses\footnote{There are several different mass conventions in the literature. For the $M_{\rm{vir}} = 10^7$M$_\odot$ models we present here, given an Einasto potential, the total mass is a factor of $\sim$3 higher, 
while $M_{300}$ is a factor of $\sim$2 lower. The mass at the half-light radius can be more than an order of magnitude 
less than $M_{\rm{vir}}$.} are difficult to constrain, the galaxies studied in these two works have (baryon$+$dark) masses of the order of $10^5-10^7$~M$_\odot$ within the half-light radius, suggesting that their virial masses range from slightly 
lower than the $M_{\rm{vir}} = 10^7$~M$_\odot$ models we present here, to significantly higher.

\citet{tolstoy09} emphasize that dwarf galaxies are not special systems and that the only reason for classifying them separately is to study
galaxy formation and evolution on a smaller scale. However, in the early universe, there are two effects that can prevent galaxies from forming or surviving in low-mass dark matter halos. The first is the epoch of reionization, when the light from the 
first stars photoionized much of the neutral gas in 
the early universe, such that neutral gas could not cool and settle in small halos. The usual assumption is that the threshold halo mass exceeds $\sim 10^8$~M$_\odot$ \citep{rees86,barkana99,gnedin00,okamoto08}.
 However, as noted in \citet{bland14}
 (referred to as Paper I), the timescale for the evaporation 
of gas is long enough that the star-forming gas in the inner region is protected and can form stars while the gas in the outer regions evaporates. Recent simulations support
this view, finding that at least some halos with masses $\sim10^7$~M$_\odot$ can survive the epoch of reionization and continue forming stars 
\citep{bovill09, ricotti09, bovill11a}.

The second process that can remove neutral gas from galaxies is the energy output of massive stars. In the two-dimensional axisymmetric models
of \citet{maclow99}, the energy from central SN explosions was found to couple efficiently with the interstellar medium (ISM) and all
the neutral gas was blown out from systems with $\leq 10^8$~M$_\odot$. \citet{maclow99} assumed a constant rate of energy input, with the lowest rate
corresponding to an SN every 3~Myr. 

In Paper I, we presented three-dimensional hydrodynamical models demonstrating 
that systems with virial masses of $10^{6.5}-10^7$~M$_\odot$ retain nearly all of their gas in the face of wind,
pre-ionization and SN from a 
$25$~M$_\odot$ star. This suggests that gas survival is common in these systems, because under typically assumed initial mass functions \citep[IMF; e.g.][]{kroupa01}, 80\% of stars massive
enough to end their lives as SNe are less massive than 25~M$_\odot$. The most important difference between the models in Paper I and previous works is that a single SN is assumed. In systems as small
as these, the low star formation rate results not only in a low SN rate, but also a high degree of stochasticity such that long gaps between SNe are
plausible. This means that the gas in some systems has sufficient time to recover from the effects of one SN before it is disrupted by another. Furthermore, an SN
event temporarily suppresses star formation such that massive stars, with their associated winds and SNe, are less likely to form in the period and location in which the gas is disrupted. 
Halos with low masses are common in the early universe and while many of the systems we study will lose nearly all of their star-forming gas before a significant amount of
stars can form, the stochasticity of star formation is such that at least some systems will have their massive stars spaced far enough apart in space and time for dense gas to
be retained and star formation to continue.

Our work also differs from previous works in that we assume an {\it inhomogenous} medium, which assists in the survival of systems as energy can escape through low-density channels. 
Paper I also investigates the effect of the location of the SN, finding that if the SN occurred
away from the center, the impact of the SN was smaller because most of the energy escaped through the lower density regions away from the center. Finally, we 
included the effects of the early-phase photoionization of the star before it explodes, finding that the energy output from a 25~M$_\odot$ star over its lifetime is comparable to the SN's explosive energy.

Our models form after the first generation of stars, but before or at the early stages of the epoch of reionization. 
They are considered to form in isolation and we do not seek to model their interaction with the Galaxy. 
This paper builds on the work of Paper I by considering what happens after the first SN. Using a probabilistic method based on that of \citet{argast00} and the
output from the simulations described in Paper I, we 
simulate star formation and chemical evolution in galaxies with masses of $10^7$~M$_\odot$. We also discuss $10^{6.5}$~M$_\odot$ systems and find that a 25~M$_\odot$ star is
sufficient to terminate star formation. We compare our results for [$\alpha$/Fe] versus [Fe/H] 
to the UFDs from \citet{vargas13} as well as Segue 2, which is the least massive known galaxy \citep{kirby13}.

Section 2 discusses the numerical models and simulations performed, although we will refer to Paper I for a more detailed explanation of the single SN simulations. 
Section 3 will outline our star formation prescription. Section 4 analyzes the distribution of mass, location and frequency of SNe, with the results of extending the
simulation beyond the first SN discussed in Section 5. Section 6 will compare to observed UFDs, followed by our conclusions in Section 7.

\section{Methods}
\subsection{Single Supernova Simulations}

We do not seek here to model the formation of the first stars in minihalos with masses of $~10^5$~M$_\odot$, 
which is dealt with in other works 
\citep{abel02, yoshida08}, but instead assume that the first stars have enriched the intergalactic medium (IGM) to [Fe/H] = $-4$ \citep{madau01,bromm04}, the lower limit of the available gas cooling functions \citep{sutherland93}.
Evidence for such low metallicities originally came from the metal-poor stars in the Galactic halo \citep{karlsson13}.
But gas phase metallicities as low as [Fe/H] = $-4$ have now been observed in the IGM at to $z\sim3-3.5$ \citep{fumagalli11}. This is therefore an appropriate starting metallicity for our model, although the threshold 
could be lower if early galaxy winds fail to enrich the IGM as in the simulations of \citet{muratov13}. [Fe/H] = $-4$ is 
close to the limit at which sufficient dust can form for low-mass star formation to be possible. We therefore allow only high-mass stars to form in unenriched gas,
with low-mass stars beginning to form in gas that is enriched even slightly from its initial state. Subsequent star formation is modeled by the prescription described in Section 3.

The simulations are described in detail in Paper I, but here we provide an explanation of the features of the first star, both during its lifetime and once it explodes. For simplicity, throughout the paper, the $10^7$~M$_\odot$ model will be referred to as M70 and the $10^{6.5}$~M$_\odot$ model as M65.

\subsubsection{Modeling the Photoionization Around the SN Progenitor}

The pre-ionization of the gas by the SN precursor plays a key role in the loss 
or survival of gas after the SN explosion, particularly in low-mass dark matter
halos. In Paper I, we consider in detail the impact of an ionized H$_{\rm{II}}$ region that
forms around the progenitor star prior to the explosion.  The key parameters
of this model are the effective temperature ($T_{\rm eff}$), the luminosity ($L$) and the associated stellar atmospheres. These give the ionizing photon flux as a function of time up to the SN
event.  Transforming the grid coordinates to spherical coordinates around the star, radial summations to approximate the optical depth integrals are performed, resulting in thermal and ionization structures calculated from the {\em MAPPINGS~IV} ionization code.
These are then mapped back to Cartesian coordinates and used in the {\em Fyris} fluid computations.  It was necessary to perform this remapping for every timestep in the fluid dynamics simulation in order to track the evolution in opacity. This slowed the code by a factor of 5--10 and was therefore only performed on specific cases of the M65 and M70
where gas had been observed to survive without pre-ionization. We considered both
clumpy and smooth external media with and without gas cooling. 

\begin{figure}[htb!]
   \includegraphics[width=0.45\textwidth]{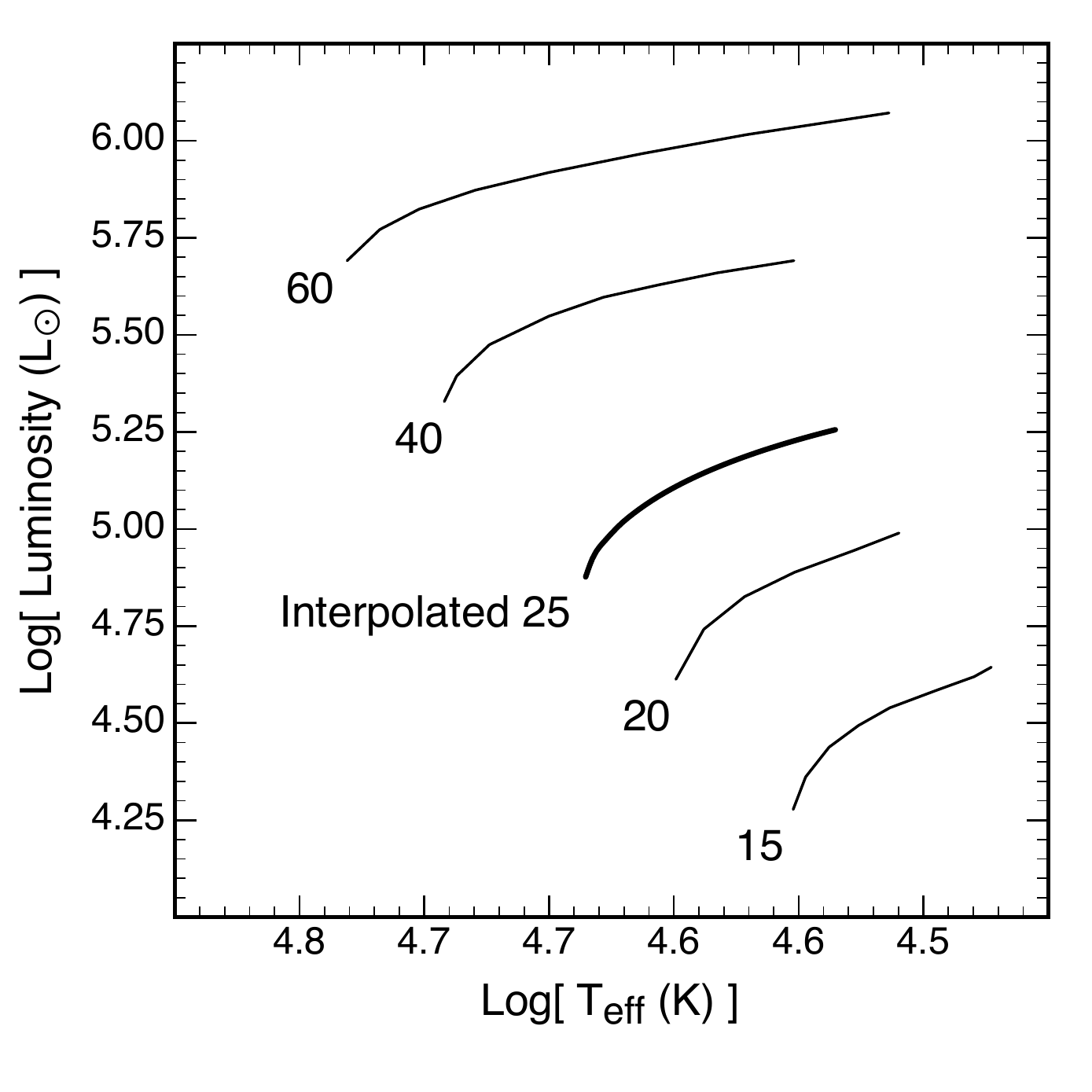} 
   \caption{Evolutionary tracks for the four most massive rotating (300 km s$^{-1}$) star models in MM02 at [Fe/H]~$\approx~-4$. We adopt a 25 M$_{\odot}$ star model interpolated from these tracks using cubic splines. For this star, the time span from the zero age main sequence (ZAMS) to the explosion is 6.1 Myr.}
   \label{f:m25spline}
\end{figure}

\begin{figure}[htb!]
   \includegraphics[width=0.45\textwidth]{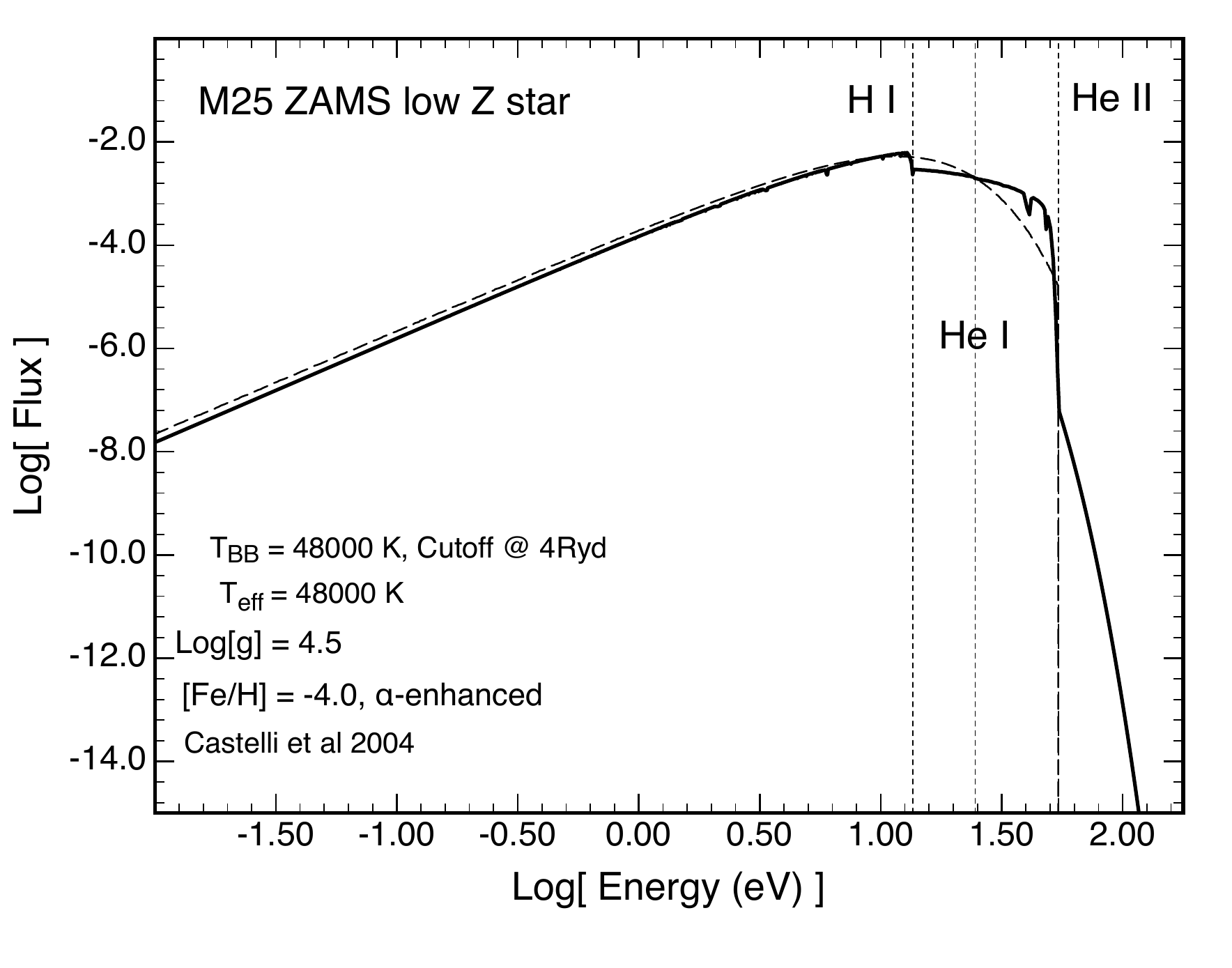} 
   \caption{Stellar spectrum (solid line) of our adopted 25 M$_{\odot}$ star
model (Figure 1) using an ATLAS9 ZAMS stellar atmosphere \citep{castelli04} at 
[Fe/H]~$\approx~-4$. The dashed line is the blackbody curve for the same surface temperature 
$T_{\rm{eff}}$ = 48 000~K.}
   \label{f:m25spec}
\end{figure}

We model the SN progenitor with a low-metallicity star of 25~M$_\odot$ since
80\% of all SN events in a conventional IMF have lower mass and therefore weaker ionizing radiation fields.
We use the evolutionary tracks of \citep[hereafter MM02]{meynet2002} for a rapidly rotating, 
low-metallicity massive star. Their [Fe/H] = $-3.7$ tracks are the closest published tracks for the metallicity threshold used here.  The 25~M$_\odot$ track was missing from that work, and so we have interpolated the main sequence phase of this model from the other stellar
models.  We used the tabulated main sequence lifetime of 6.1 Myr from MM02.  The interpolation is shown in Figure~\ref{f:m25spline}. For the above model,
we adopt a stellar spectrum at [Fe/H] = $-4$ using an ATLAS9 ZAMS stellar atmosphere 
\citep{castelli04} shown in Figure~\ref{f:m25spec}.

To avoid the complexities of rapid post-main sequence evolution, we assume that the star simply explodes as an SN at the end of the main sequence.  \citet{ekstrom2006} showed that mass loss during the main sequence in similar stars at very low-metallicity may only amount to 1\% of the initial mass, and \cite{kudritzki2002, kudritzki2005} showed that the stellar wind luminosities of low metallicity ([Fe/H]$\le -4.0$) O stars are typically $10^{34}$ erg s$^{-1}$ or less, several magnitudes below an equivalent solar metallicity star.  We combine these values to get a mean wind velocity, for a main sequence mass loss of 1\%:
\begin{equation}
\resizebox{.85\hsize}{!}{$v_w \sim  880 \left(\frac{L_w}{10^{34} {\rm erg\: s^{-1}}}\right)^{1/2}
\left(\frac{M}{25~{\rm M_\odot}}\right)^{-1/2}
\left(\frac{t_{\rm MS}}{6.1~{\rm Myr}}\right)^{-1/2} \; {\rm km\: s^{-1}}$}.
\end{equation}
This wind is included in the pre-SN evolution, but the small mass flux and low ram pressure meant that it had negligible effect on the simulations.  Once the H$_{\rm{II}}$ region pressurized the surrounding region, the stellar wind was unable to blow a wind bubble of 
any significant size.

\begin{figure}[ht!]
   \includegraphics[width=0.45\textwidth]{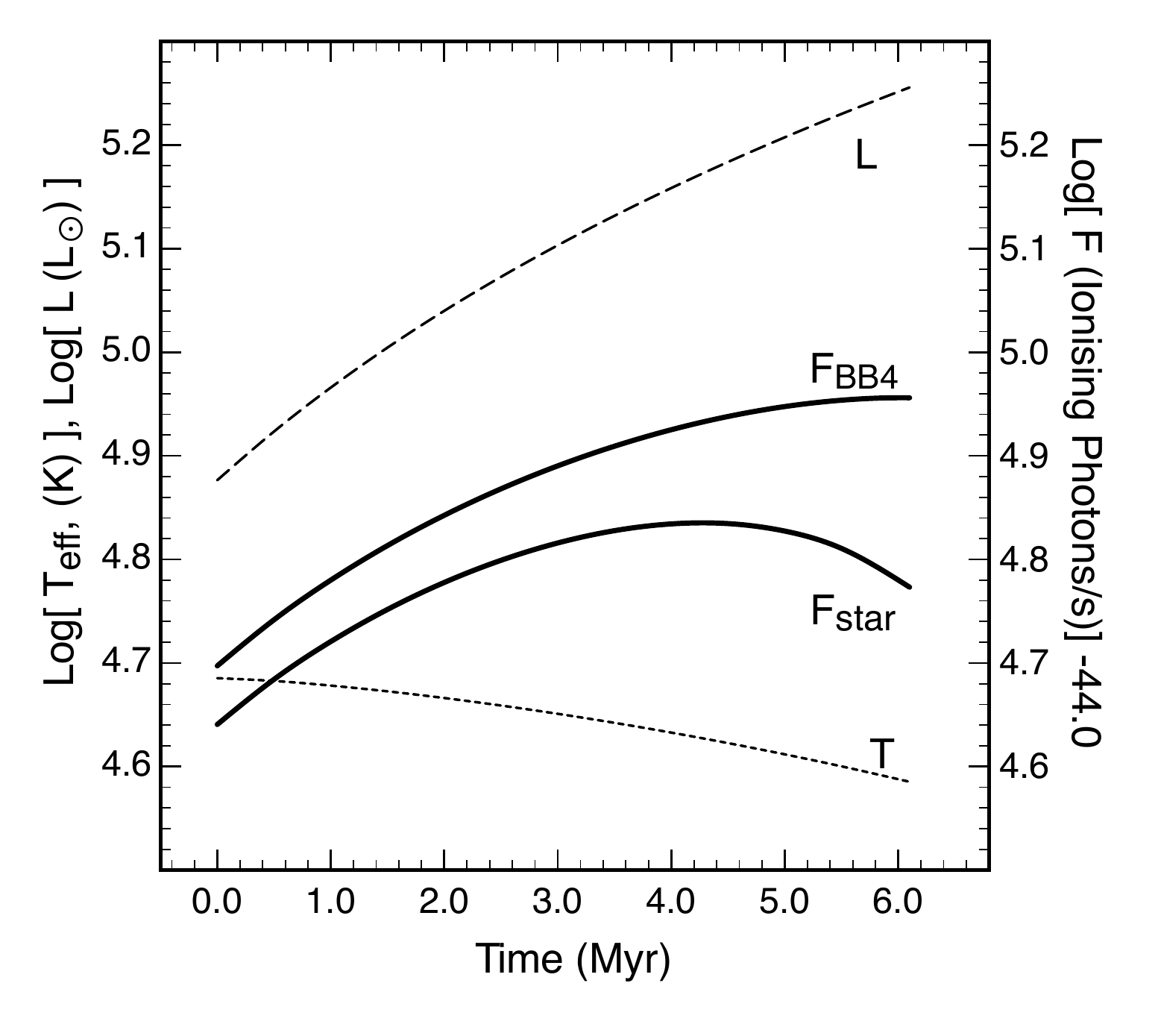}
   \caption{Temporal evolution of our adopted 25 M$_{\odot}$ star model from
the ZAMS to detonation. We show the evolution of the total luminosity
(dashed line), the effective temperature (dotted line), the
equivalent blackbody and the star model flux (upper and lower
solid lines respectively). The solid lines are the total ionizing
flux in units of 10$^{44}$ photon s$^{-1}$ (see the right-hand side).}
   \label{f:star}
\end{figure}

The key parameter for these near-zero metallicity O stars is their unusually high effective temperatures for a given mass.  At 25~M$_\odot$, we estimate the initial temperature to be 48,000~K , compared to a similar mass star at solar metallicity of only 40,000~K (see, for example \citet{meynet2000} and more recently \citet{georgy2012}).  This means the star will be more effective as a source of ionizing photons than the equivalent star in the solar neighborhood. To convert the evolutionary tracks, $T_{\rm eff}(t)$ and $L(t)$,  to an ionizing photon luminosity, we used the ATLAS9 \citep{castelli04} atmospheric grid, which uniquely contains a set of atmospheres for [Fe/H]$ = -4.0$  with alpha element abundance enhancements, perfectly matching the initial gas composition in our simulations.  
The interpolated stellar temperature, luminosity and ionizing fluxes are shown in Figure~\ref{f:star}.  For comparison, a photon flux assuming a blackbody instead of the stellar atmosphere is included.

\subsubsection{Effects of the Supernova Explosion}

\begin{figure}
     \centering
     \subfigure{
          \label{f:M70CCgas}           
          \includegraphics[width=.45\textwidth]{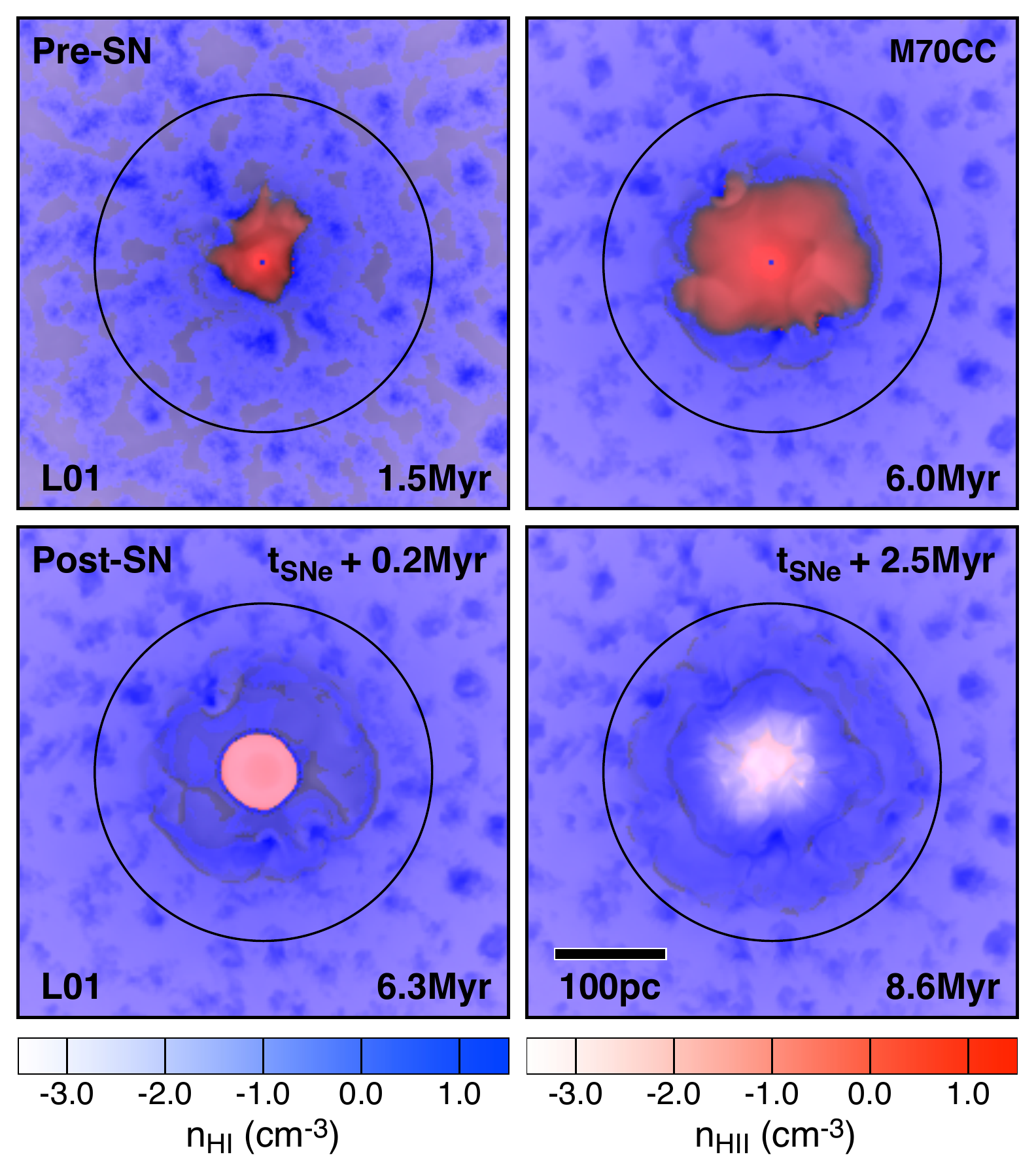}}
     \subfigure{
          \label{f:M70OCgas}
          \includegraphics[width=.45\textwidth]{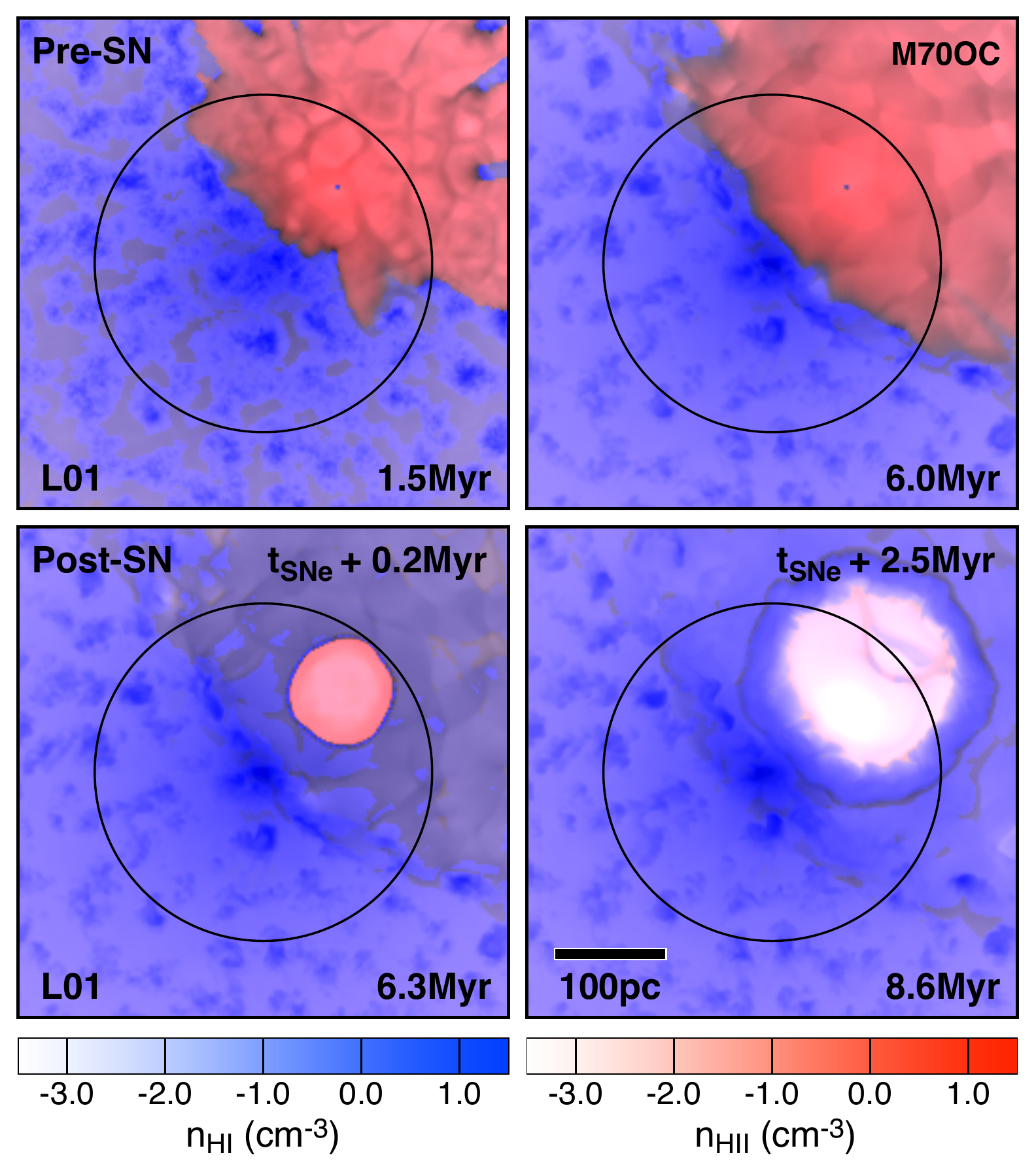}}
      \caption{Ionized and neutral gas densities during the early times of the M70 centered (top) and off-center (bottom) models. The pre-ionization phase removes the neutral gas from the surroundings, resulting in the supernova affecting a larger region. In the off-center case a significant amount of dense gas remains on the side opposite the supernova.}
     \label{fig:M70gas}
\end{figure}

The effect of the SN on the gas was studied in Paper I and \citet{bland11}, and is shown in Figure~\ref{f:M70CCgas} for the case of a centered explosion in an 
M70 system and Figure~\ref{f:M70OCgas} for the off-center case. Here we use only the models
that adopt a clumpy ISM and incorporate radiative cooling, which are more realistic. Radiative 
cooling means that only $\sim10\%$ of the energy is available for 
driving the gas, which is consistent with previous studies \citep[e.g.][]{thornton98}. The clumpiness 
of the ISM has the effect of decreasing the efficiency with which the SN
energy couples to the gas, meaning that a significant proportion of the energy escapes into the IGM 
and more gas is retained. This also reduces enrichment, as metals escape rather than being mixed in.

\begin{figure}[htb]
\begin{center}
\includegraphics[width=.45\textwidth]{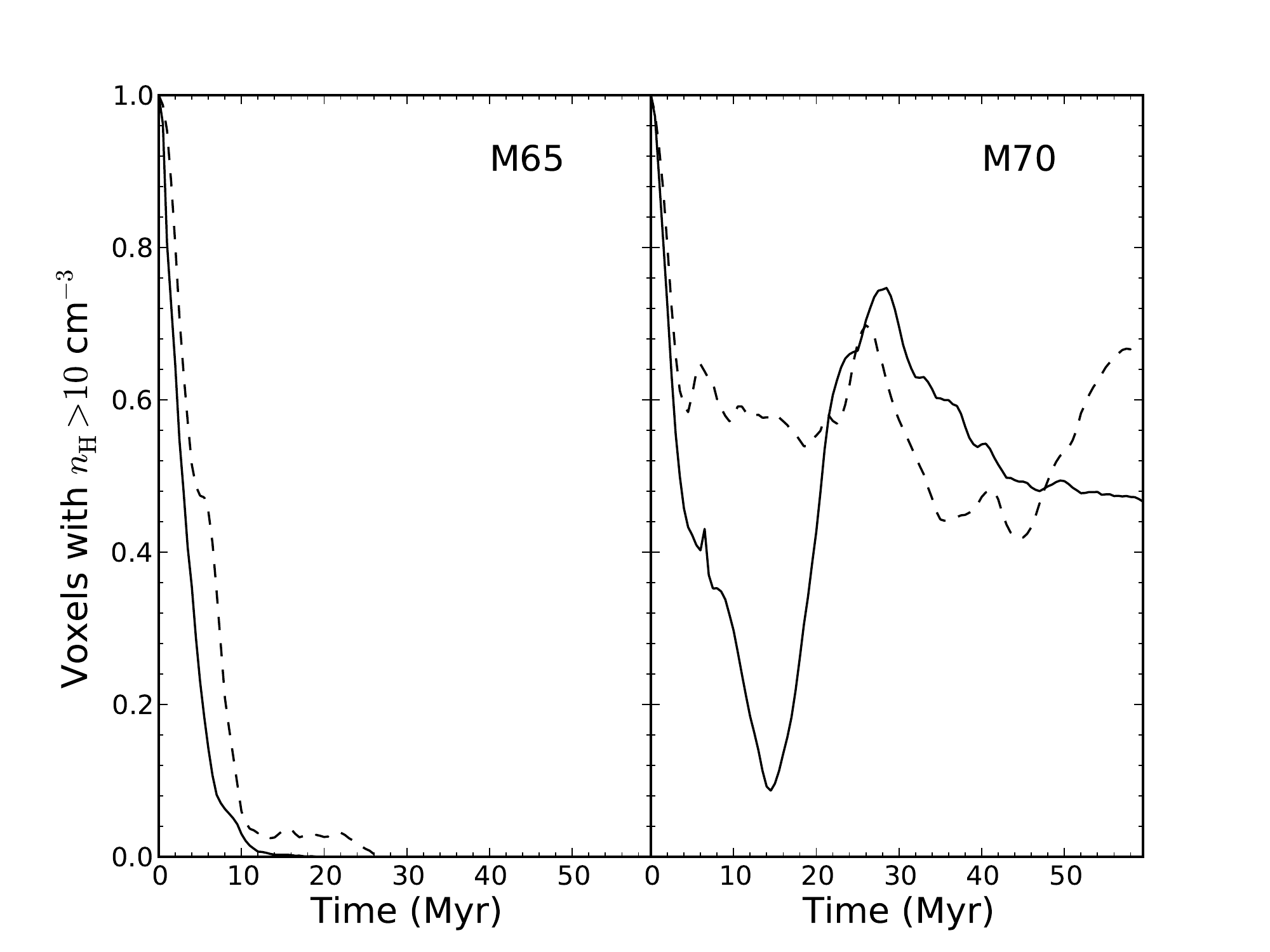}
\caption{Dense gas retention by the M65 systems (left) and M70 systems (right) for both the central (solid line) and off-center (dashed line) explosions expressed
 as a proportion of the initial number of dense cells. The M65 models lose all of their dense gas very quickly, while the M70 systems recover within 20~Myr.}
\label{f:propdens}
\end{center}
\end{figure}

In all cases, the pre-SN phase 
drives a significant proportion of the gas out of the center of the system. The models with off-center explosions are more resilient than
those where the explosion occurs in the center. This is because the gas on the opposite side of the explosion is protected by the dense gas close to the center. 
In the central case, all of the SN energy pushes on dense gas rather than much of it escaping. The M70 central model shows 
two stages of dense gas loss: one associated with the ionization from the stellar winds and a second with
the SN explosion. The M70 off-center model does not show a second dip from the SN because the small amount of dense gas pushed out
is balanced by the extra dense gas formed due to compression by the shock that wraps around the center of the galaxy. The proportion of dense gas 
retained is compared in Figure~\ref{f:propdens}. The M65 models both lose all of their dense gas due to the effects of the massive star.

\subsection{Multiple Supernovae}

The simulations outlined in Paper I followed the evolution of the system in the face of the energy output from a single massive star and it was found that for systems more
massive than $10^{6.5}$M$_\odot$, gas will be retained and further star formation is possible. The goal of this paper is to model this subsequent star formation and 
chemical enrichment, allowing the production of [$\alpha$/Fe] versus [Fe/H] plots to compare with observed UFDs. The simulation method described in the previous section is 
computationally intensive, and the boundary conditions begin to have an effect even at 60~Myr. A significantly larger grid would be required to run the full simulation for
hundreds of Myr. Instead, we use the 60~Myr of data from the first SN to simulate the effects of later SNe.

We run a simulation for 600 Myr with a resolution of 0.5~Myr. Stars are allowed to form using a Kroupa IMF and an adaptation of the \citet{argast00} 
stochastic method, where stars
are formed in a cell with a probability proportional to the square of the gas density as described in Section 3. The gas recovers within 30~Myr of an SN explosion from a 
25~M$_\odot$ star, so the density in each cell of 
the single SN simulations from Paper I at 35~Myr is taken as representing the undisturbed density. 
If a star more massive than 20~M$_\odot$ is formed sufficiently close to the center to have a significant impact on the
system, the density is reset to the $t~=~0$ value and allowed to evolve
forward through the frames of the single SN simulation. If a star with a mass between 8 and 20~M$_\odot$  is formed, the simulation keeps moving forward through the
frames during the lifetime of the star, then resets to the 6 Myr state, which is the time at which the SN occurs in the single-SN simulation. 
This has the effect of removing the pre-ionization phase, which is less significant for stars less than 20~M$_\odot$.
Note that the impact of an SN will be lessened in the absence of a pre-ionization phase, so this may slightly underestimate the amount of star formation for a few Myr
after the SN. Opposing this effect is the fact that the gas will not always recover completely before the next SN occurs, so a SN from a less massive star
will sometimes occur in a pre-ionized environment.

\begin{figure}
\begin{center}
\includegraphics[width=.45\textwidth]{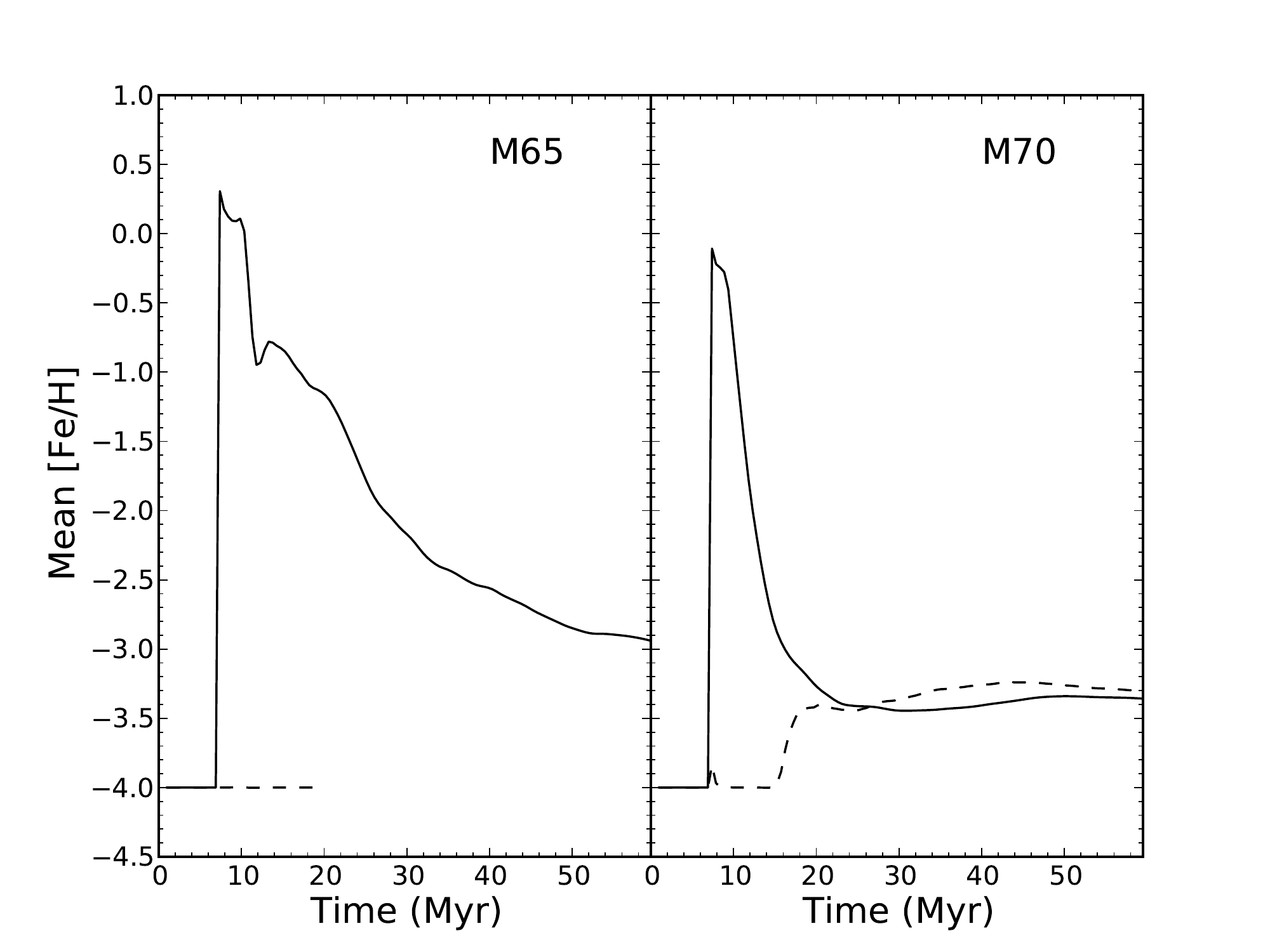}
\caption{Average [Fe/H] in the central region ($r < 50$~pc, solid) and in the densest cells ($n_{\rm{H}} > 10$, dashed) for the M65 and M70 central models. In the M65 models
no dense cells remain after 20~Myr. While the metallicity in the central region is higher than for the M70 case, the density is too low for stars to be formed.
 In the M70 model a few stars 
form at the high metallicities in the central cells before the 
metals are mixed into the gas, but more will form at the less enhanced metallicity in the dense cells. The mean metallicity is close to constant after 30~Myr.}
\label{f:meanfe}
\end{center}
\end{figure}

A similar method is used to calculate [Fe/H], except that instead of resetting as is done for density, each SN enriches the gas, so [Fe/H] increases.
As can be seen in Figure~\ref{f:meanfe}, by 40~Myr after the SN, the [Fe/H] in the densest cells and the central cells is close to constant, so this value can be used to calculate
the iron output of a single SN. Models of Type II SN yields such as in \citet{nomoto06} note that the iron yield from Type II SNe is 
nearly independent of stellar mass. Therefore, after each SN,
we reset the [Fe/H] as was done for density above, but with the addition of the iron from a single SN multiplied by the number of previous SNe. This is only an approximation and the true Fe yield may be slightly higher or lower depending on the state of the gas as was noted for density above. [$\alpha$/Fe]
 is calculated in the same way as [Fe/H], with the $\alpha$ abundance defined as the sum of the abundances of Ca, Mg, Si and Ti.  
 The yields used are from \citet{woosley95}, interpolated and extended to 8~M$_\odot$. A small amount of noise is added
to the yields in each cell to account for SN asymmetry and unresolved properties of the gas. The behavior predicted in \citet{frebel12} that [Fe/H] should show scatter of 
$\sim1$~dex, but that there should only be small scatter in [$\alpha$/Fe], is shown in our results.

Type Ia SNe commence at approximately 100~Myr and can therefore be used as a test of whether a system has formed stars for longer than this \citep{frebel12, 
vargas13}. We adopt the \citet{jiminez14} Type Ia rates scaled down to the star formation rates of our systems, suggesting a Type Ia SN every 15~Myr. The Type Ia SN yields used are from \citet{iwamoto99}, with 0.72~M$_\odot$ of Fe and 0.339~M$_\odot$ of alpha elements 
ejected. This is seven times the amount of iron that a typical Type II SN produces and it therefore only takes a few Type Ia SNe to significantly reduce [$\alpha$/Fe].

\section{Star Formation Prescription}

A realistic prescription of star formation is required to predict the observational properties of our simulated systems. The amount and location of star formation are the most
important factors in the length of time stars will form before there are multiple SNe close enough in location and time to turn off star formation. Particularly important is whether the system survives for the 100~Myr it takes for the first Type Ia SNe to occur, 
which shows up observationally in decreased [$\alpha$/Fe]. 

The first step was to choose the criteria for a cell in the simulation to form a star. 
The simulations do not include self-gravity and molecular cooling so the gas does not cool to star-forming temperature,
 which is much less than 100~K. Recent work by \citet{hopkins13} outlines several possible methods for specifying
 the amount and position of stars in parsec-resolution simulations. They find that the best results are given by an `overdensity' criterion:

\begin{equation}
\alpha = \beta \frac{|\nabla \cdot \bf{v}|^2 + |\nabla \times \bf{v}|^2}{G\rho} < 1
\end{equation}

where the density of the gas $\rho$ is compared to the sum of the square magnitudes of the divergence and curl of the velocity field $\bf{v}$, which account for
the local velocity dispersion and the internal rotation and shear of the gas. $\beta \sim 0.5$ is a constant relating to geometry. 

However, \citet{hopkins13} note that for systems where the average density is much 
less than the typical densities at which stars form, a simple density criterion works nearly as well. This is certainly the case for our modeled systems, and a test of
both criteria over 10~Myr adjusted for a chosen star formation rate did not show significant differences in the spatial distribution of star formation.

The density criterion is simpler than the Hopkins overdensity criterion and we therefore adapt the method of \citet{argast00}. At each time step, $5\times 10^4$ cells 
are selected at random from the $216^3$ L1 grid, and each is given a probability of forming a star that is proportional to the square of the density of the cell. 
This introduces stochasticity to the method, ensuring
 that the star formation is not completely dominated by the cells in the center, which have the highest density. Changing the proportionality coefficient
 and the number of cells gives two free parameters that allow the rate of star formation and the distribution of densities at which stars form to be specified.

The critical metallicity at which the IMF switches from top-heavy to bottom-heavy is believed to be 
$Z_{\rm{crit}} \sim 10^{-6}-10^{-3.5}$~Z$_\odot$ \citep{bromm11}. Given that the metallicity in the first galaxies is likely to be similar to the upper
 end of this range \citep{bromm03,frebel07}, we allow low-mass stars to form only in gas that has been enriched by our first SN. 
We note that some simulations \citep{johnson08,greif10} find that the metallicity of the first galaxies could be as high as [Fe/H] = $-3$, which would have the effect
of increasing our total star formation.
The masses of stars are selected by sampling a Kroupa IMF \citep{kroupa01} with a mass range of $0.1-50$~M$_\odot$, with the restriction that if the mass
of the star is less than 8~M$_\odot$ and the metallicity at the location the star is to be formed is [Fe/H] $\leq -4$, the star does not form. 

We start with a reference model to find the baseline star formation rate, allowing us to calculate the disrupting effect of the SN, 
as well as providing a check that our star formation criteria are reasonable. This is the star formation rate for the undisturbed gas if no SNe occurred, assuming
the Kroupa IMF at all metallicities. The mean star formation rate is $1.3\times 10^{-5}$~M$_\odot$yr$^{-1}$ for M70 and $5.7\times 10^{-6}$~M$_\odot$yr$^{-1}$ for M65. 
This corresponds to $1.4\times 10^{-4}$~M$_\odot$yr$^{-1}$kpc$^{-2}$ within the scale radius of M70 and $1.8\times 10^{-4}$~M$_\odot$yr$^{-1}$kpc$^{-2}$ for M65, 
which is within an order of magnitude of the observed star formation rates in starbursts in larger dwarf galaxies such as Carina \citep{bigiel08}.

\begin{figure}
     \centering
     \subfigure{
          \label{f:metalsM70CC}           
          \includegraphics[width=.45\textwidth]{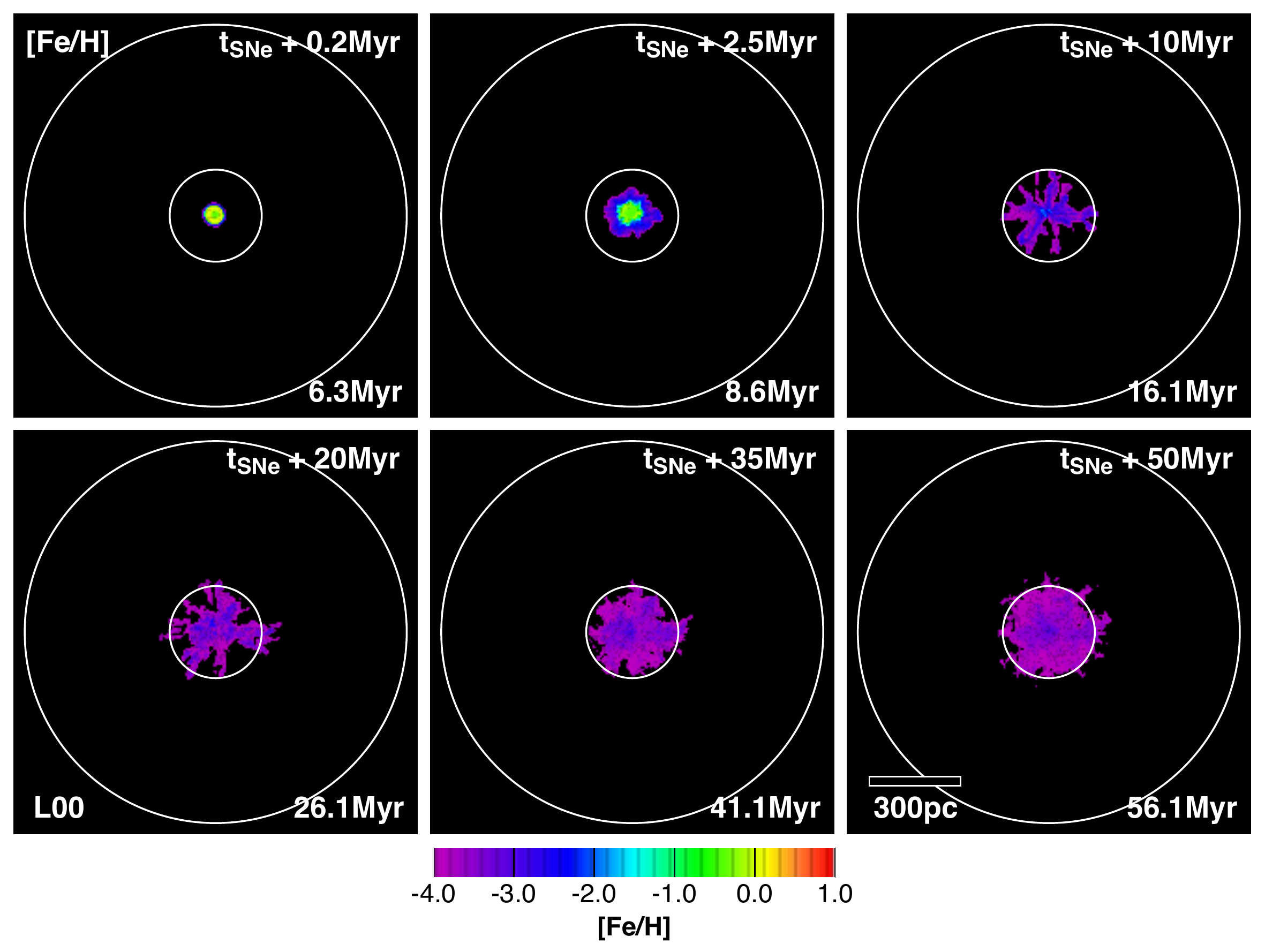}}\\
     \subfigure{
          \label{f:metalsM70OC}
          \includegraphics[width=.45\textwidth]{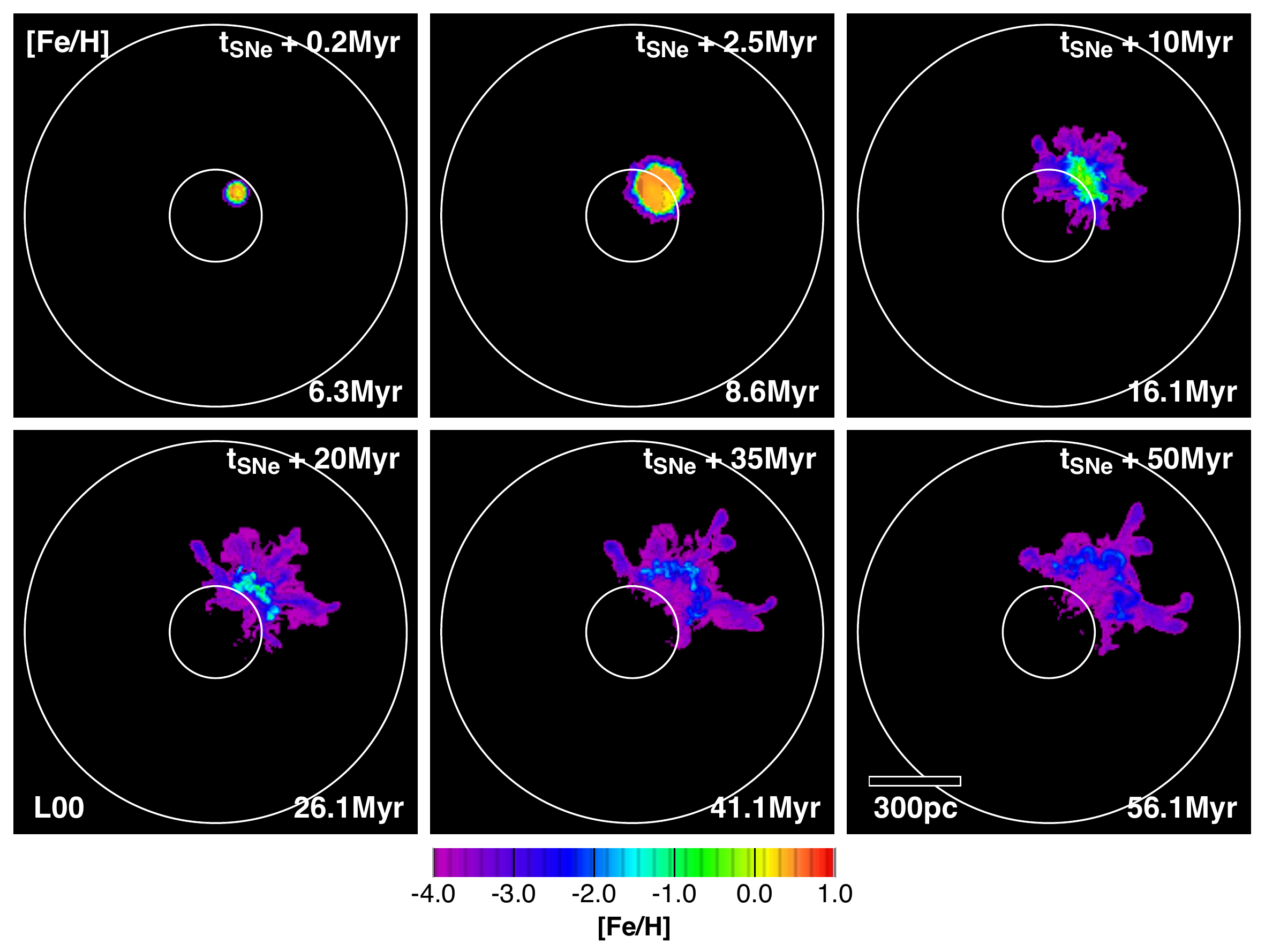}}
      \caption{Enrichment of the system over time in the M70 central case (top) and the off-center case (bottom). The dense gas is enriched only in the central case. }
     \label{f:metalsM70}
\end{figure}

We then take the single-SN model and assume that a 25~M$_\odot$ star forms at $t = 0$. The proportion of dense gas 
retained is compared in Figure~\ref{f:propdens}, while Figure~\ref{f:metalsM70} shows how the distribution of 
metals changes over time after the SN. In the M70CC case, the metals mix with the gas in the center and the regions with dense gas become enriched. In the
M70OC case, the gas in the center remains unenriched and therefore no low-mass stars are formed. An SN explosion closer to the center 
is therefore required for low-mass stars to form.

\begin{figure}[htb]
\begin{center}
\includegraphics[width=0.45\textwidth]{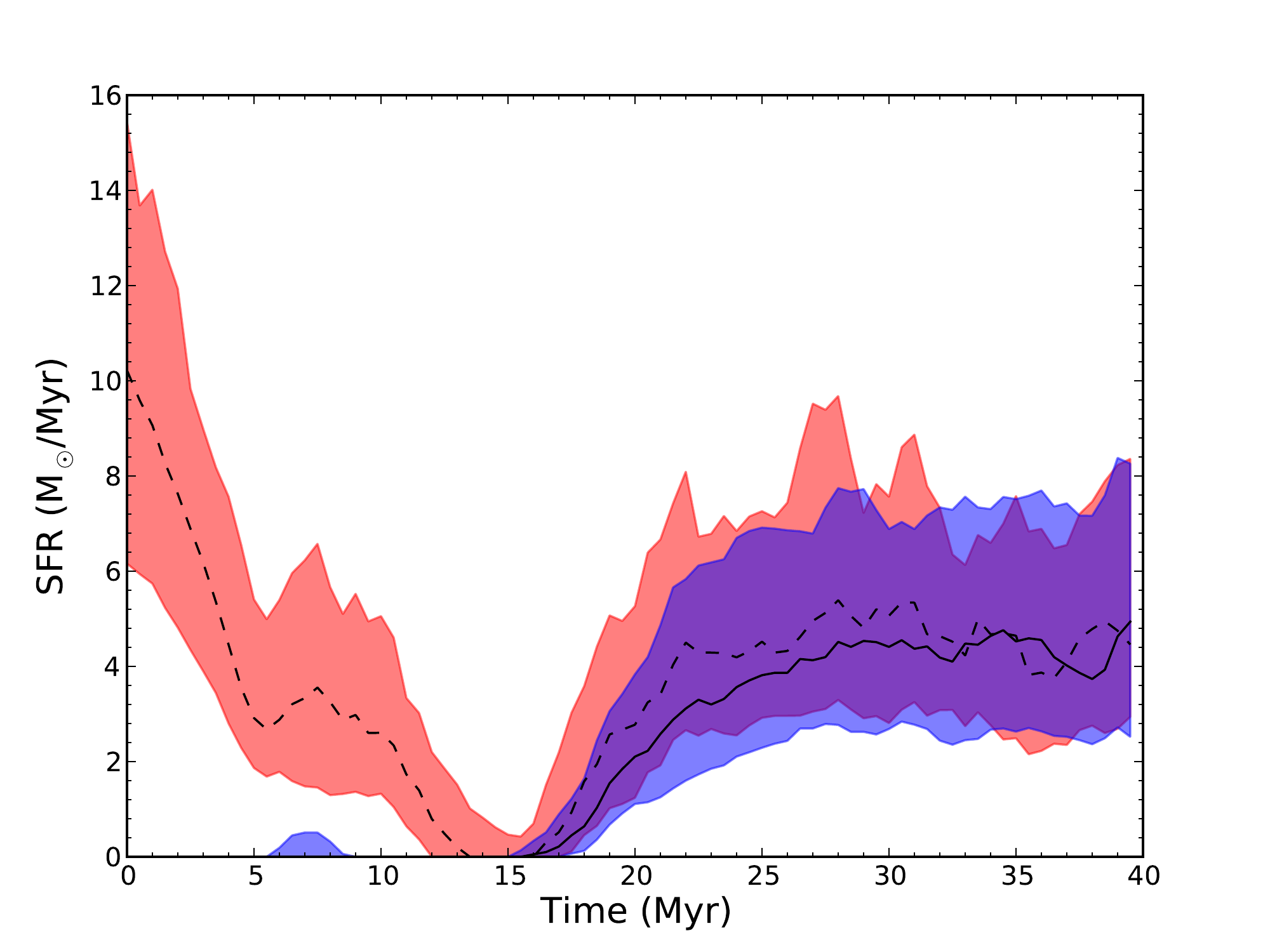}
\caption{Total star formation integrated over the M70CC galaxy simulated over 100 runs, where stars with $M < 8$~M$_\odot$ can form at any metallicity 
(dotted line is median, red region is the interquartile range) and where low-mass stars can only form at [Fe/H] $> -4$ (solid line is median, blue region is interquartile range). The two converge very quickly, suggesting that by the time the gas recovers sufficiently for further stars to form, at least some of the metals have been mixed into it.}
\label{f:M70CCSFR}
\end{center}
\end{figure}

The variation in star formation rate given the wind and SN from this 25~M$_\odot$ star is shown in Figure~\ref{f:M70CCSFR} for M70CC. 
The star formation rate dips 
as much of the dense gas is blown out by the energy from the stellar winds and SN, but eventually much of the 
gas returns to the center of the galaxy, meaning that there are enough dense regions for the star formation rate to recover to 50\% of the pre-SN level. The M70 
off-center explosion does not form low-mass stars because the dense gas is not enriched.  The M65 models quickly lose all of their dense gas, so they never form low-mass stars. 

In the M65 model with a central explosion, dense gas is rare by the time the SN occurs and the SN blows out the remaining $10\%$. 
By 15~Myr, the amount of gas denser than $n_{\rm{H}} =$~10~cm$^{-3}$ is negligible. The off-centred M65 model survives the pre-sN ionization with about half its 
dense gas intact. However, the SN explosion blows out the rest of the dense gas and by 25~Myr a negligible amount of gas remains with $n_{\rm{H}} > 10~$cm$^{-3}$. 
The mean total star formation with the enrichment condition relaxed is 16~M$_\odot$ for the central model, while the mean for the off-center model is 27~M$_\odot$.

In this section we discussed only the impact of the first high-mass star and found that for the M70 model, after the first central explosion, low-mass stars are able to form, while an
off-center explosion has only a small impact on the system. This means that the system will survive and form stars until there are multiple SNe close to the 
center and not spaced too far apart in time. In the next section, we discuss the distribution of the masses, location and timing of subsequent stars that will end their life as an SN, all of which affect how long star formation continues in the system.

\section{Mass, Location and Timing of Supernovae}

For the multi-SN simulation, we require estimates of how frequently massive stars form, how massive they are, and where they occur. Here we use the output of the
single SN simulations to estimate these parameters. We also identify effects that separate massive stars in location and in time, supporting the idea that our systems
can form stars for 100~Myr and beyond. It has already been noted that the M65 systems cannot survive the effects of the 25~M$_\odot$ star, so here we focus on M70 only.

\subsection{Masses}

The mass distribution of stars is given by a Kroupa IMF with a lower bound of 0.1~M$_\odot$ and an upper bound of 50~M$_\odot$. The proportion of stars 
with a mass greater than $M$ is therefore

\begin{equation}
p(m > M) = \frac{\int_M^{50} m^{-\alpha} dm}{\int_{0.1}^{50} m^{-\alpha} dm},  
\end{equation}
where for the Kroupa mass function $\alpha = 2.3$ for $M > 0.5$ and $\alpha = 1.3$ for $M < 0.5$. From this, we find that $0.6\%$ of stars have a mass 
greater than $8~$M$_\odot$, which we take to be the threshold mass at which a star will end its life as an SN. 
Table 1 shows the percentage of SN-forming stars that are in each mass range. The median SN-forming star has a mass of $\sim13~$M$_\odot$ and only 16\% 
are more massive than the 25~M$_\odot$ star we use in our simulations. Table 1 also gives the lifetimes and the photon flux for various 
stellar masses from \citet{schaerer02} 
for $Z = 0$ and $Z = 0.02$~Z$_\odot$. At lower metallicity the lifetimes are shorter. The starting metallicity in our simulations is $10^{-4}~$Z$_\odot$, so the 
lifetimes and Q in our simulations will be somewhere between these two extremes.   

\begin{table}
\caption{Properties of Stars as a Function of Mass}
\begin{tabular}{c c c c c c}
\hline \hline\\
& &\multicolumn{2}{c}{Z=1/50~Z$_\odot$}&\multicolumn{2}{c}{Z=0}\\
\hline
Mass&Prob&Lifetime&$\bar{Q}$&Lifetime&$\bar{Q}$\\
(M$_\odot$)& &(Myr) & (photon s$^{-1}$)&(Myr)&(photon s$^{-1}$)\\
8-10&0.28&38--27&0.2--1.5$\times 10^{46}$&25--18&2.3--4.7$\times 10^{47}$ \\
10-12&0.17&27--21&1.5--6.5$\times 10^{46}$&18--14&4.7--8.5$\times 10^{47}$\\
12-15&0.16&21--15&0.7--3.0$\times 10^{47}$&14--10&0.9--1.7$\times 10^{48}$\\
15-20&0.15&15--11&0.3--1.4$\times 10^{48}$&10--7.5&1.7-3.8$\times 10^{48}$\\
20-25&0.084&11--8.3&1.4--3.9$\times 10^{48}$&7.5--6.1&3.8-6.8$\times 10^{48}$\\
25-30&0.053&8.3--7.0&3.9--7.6$\times 10^{48}$&6.1--5.2&0.7--1.1$\times 10^{49}$\\
30-35&0.036&7.0--6.1&0.8--1.2$\times 10^{49}$&5.2--4.6&1.1--1.6$\times 10^{49}$\\
35-40&0.026&6.1--5.5&1.2--1.8$\times 10^{49}$&4.6--4.2&1.6--2.1$\times 10^{49}$\\
40-45&0.019&5.5--5.0&1.8--2.3$\times 10^{49}$&4.2--3.9&2.1--2.8$\times 10^{49}$\\
45-50&0.015&5.0--4.7&2.3--2.9$\times 10^{49}$ &3.9--3.7&2.8--3.4$\times 10^{49}$\\
\hline
\end{tabular}

\end{table}

\subsection{Timing}
The probability p$_{\rm{form}}$ that at least one star with a mass greater than $8$~M$_\odot$
 has formed since $t = 0$ is

\begin{equation}
p_{\rm{form}}(t,M>M_{\rm{crit}}) = 1-0.994^n
\end{equation}

where $n$ is the number of stars that have formed by time $t$, $M_{\rm{crit}} = 8~$M$_\odot$ is the threshold for a star to end its life as an SN, and the probability $p_{\rm{SN}}$ that an SN explosion has occurred before time $t$
is:

\begin{equation}
p_{\rm{SN}}(t) = 1 - \prod_{M=M_{\rm{crit}}}^{50} (1-p_{\rm{M}}(t - t_{\rm{M}}))
\end{equation}

where $t_{\rm{M}}$ is the lifetime of a star of mass $M$ and $p_{\rm{M}}$ is given by Equation 2.

\begin{figure}[htb]
\begin{center}
\includegraphics[width=.45\textwidth]{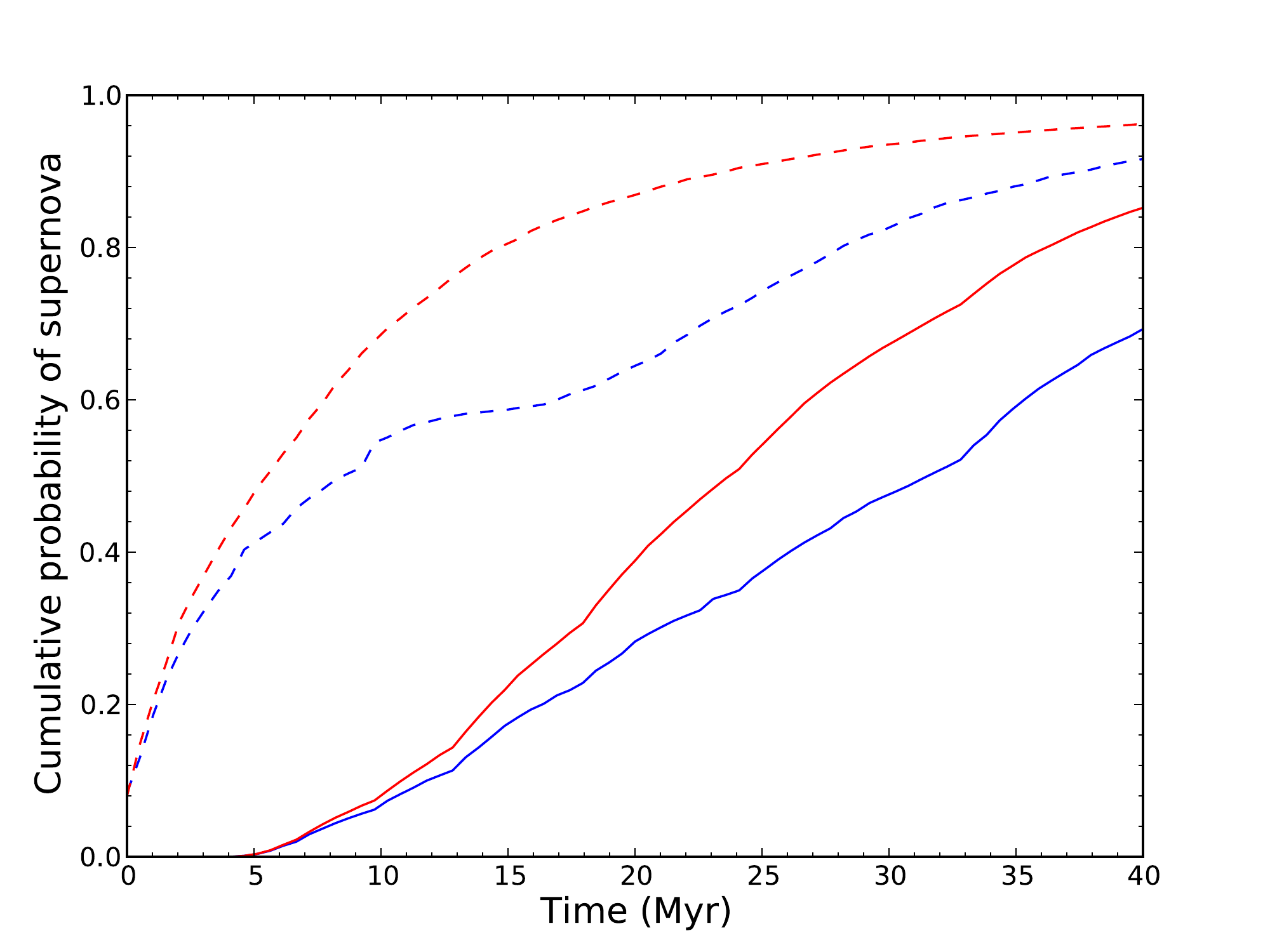} 
\caption{Probability of a second supernova-forming star having formed (dashed line) and exploded as a supernova (solid line) as a function of time after the 25~M$_\odot$ star formed at $t=0$ for the central model (blue) and the off-center model (red). This shows that a second supernova is unlikely to form before the gas has recovered from the first, meaning that in the majority of cases star formation 
will continue for at least 50~Myr.}
\label{f:M70probSN}
\end{center}
\end{figure}

Figure~\ref{f:M70probSN} shows the cumulative probability of a second SN with time following the formation of a 25~M$_\odot$ 
star at $t=0$. For the centered case, while a second massive star 
is likely to form within $10$~Myr, there is only a $\sim$30\% chance that the SN from the star will occur before the gas recovers at 
$t\sim20$~Myr. For the second explosion to occur within $20$~Myr of the first, the second star must have a mass of at least $10-12~$~M$_\odot$. This supports our assumption that the gas usually recovers after an SN before the next SN disrupts it again, meaning that the system 
is likely to survive at least its first two SN explosions.

\subsection{Location}

Figures~\ref{f:M70CCgas} \& \ref{f:M70OCgas} show the effects of a 25~M$_\odot$ star on an M70 halo. Over the first 6~Myr, the star pushes gas out as well as 
ionizing the surrounding 
gas, creating a region where there is no dense neutral gas for star formation to occur. This region forms a Str{\"o}mgren sphere and has a radius:

\begin{equation}
r_{\rm{strom}} = (\frac{3Q(H)}{4\pi\beta_2n^2})^{1/3}
\end{equation} 

where $Q(H)$ is the photon flux from the star and $n$ is the gas density. At low metallicities, the recombination rate $\beta_2$ to all levels other than the ground state is
$\beta_2$ = $1.84\times 10^{-13}\times (T/1.5\times 10^{-4}~$K$)^{-0.78}$ \citep{dopita03}. 

\begin{figure}
     \centering
     \subfigure{
          \label{f:strom25}           
          \includegraphics[width=.45\textwidth]{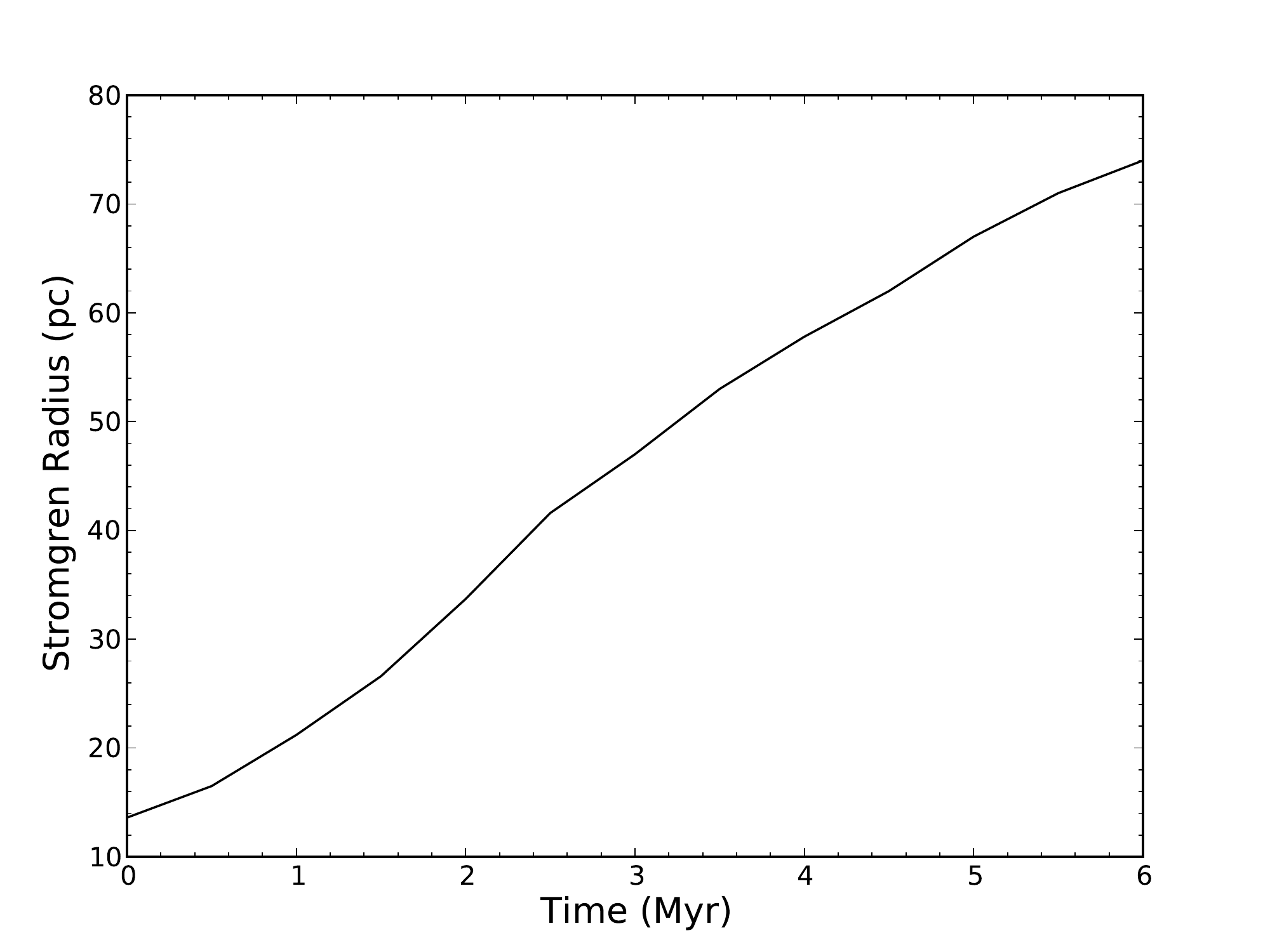}}
     \subfigure{
          \label{f:strominit}
          \includegraphics[width=.45\textwidth]{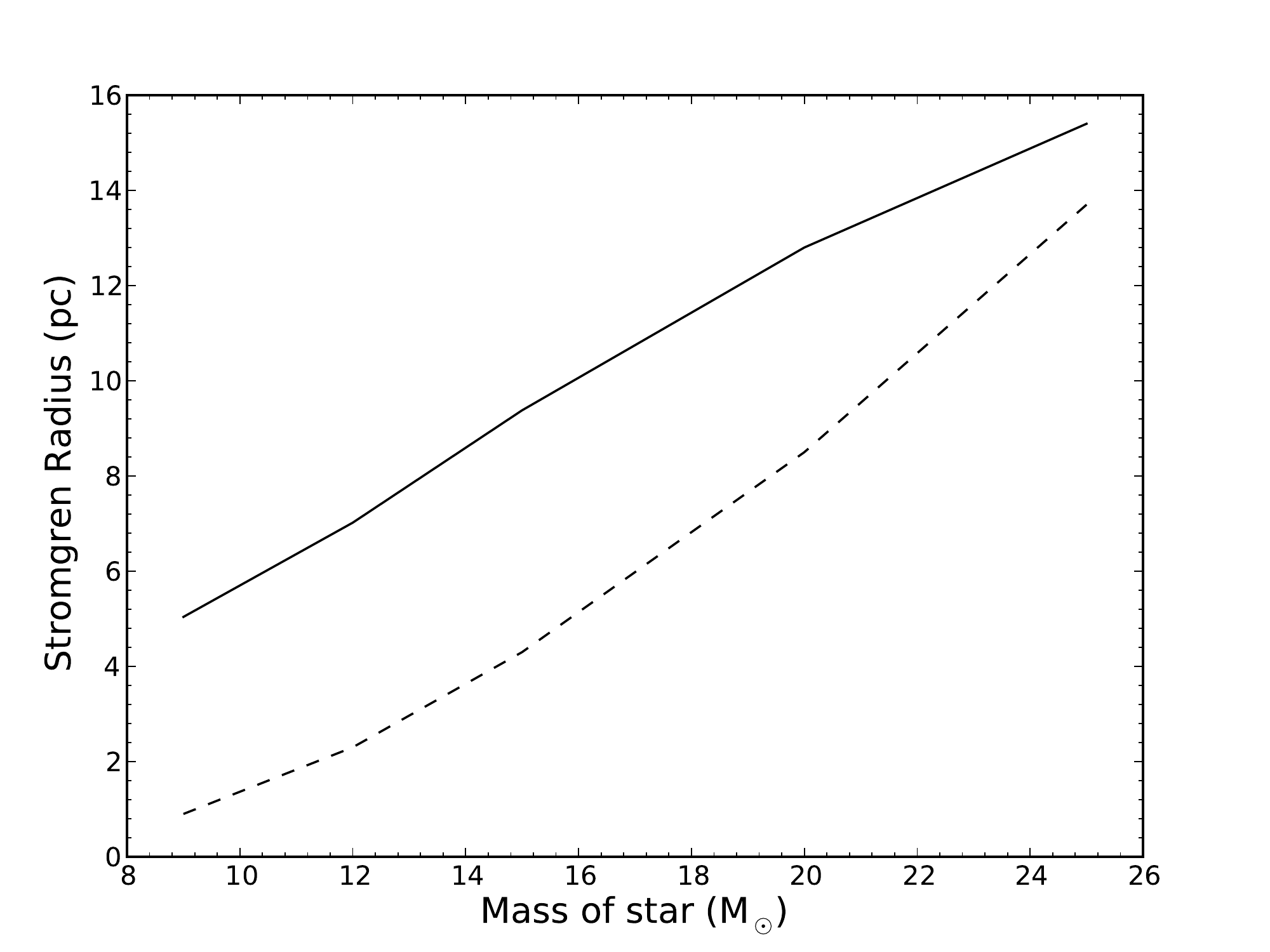}}
      \caption{Top: the Str{\"o}mgren radius of a 25~M$_\odot$ star as a function of time. The Str{\"o}mgren radius increases with time as the star removes much of the neutral gas from its environment. This creates an exclusion zone where star formation is temporarily stopped. Bottom: the Str{\"o}mgren radius of a star in the initial gas density profile as a function of mass for the case of zero metallicity (dashed line) and [Fe/H] = $-1.7$ (solid line). The affected volume rises as a steep function of stellar mass, with the more common supernova-forming stars having a very small Str{\"o}mgren radius and therefore a pre-ionization phase that can be neglected.}
     \label{fig:strom}
\end{figure}

The gas density is not constant, but decreases as radius increases and is also inhomogenous such that it is not constant at a given radius. In the case where the star
is not at the center, this means that the ionized region can be highly asymmetric, as the Str{\"o}mgren radius is smaller in the direction of the dense gas in the center and
much larger in other directions. 

In the central case, the region is nearly symmetrical and we use the mean density of the gas. The extent of the Str{\"o}mgren sphere increases with time, because the 
region surrounding the star becomes less dense. Figure~\ref{f:strom25} 
shows how the radius increases for a 25~M$_\odot$ star at the center of 
the M70 system. The values of temperature and photon flux are from MM02 as described in Section 2.1.1.
This Str{\"o}mgren radius can be considered an exclusion radius within which no stars will form. The maximum radius of 74~pc at 6~Myr is very close to the half gas-mass
radius where we form the star in the off-center model.

The above effect can aid the survival of small galaxies. Two massive stars cannot form close together in location, as the first will create an ionized region that prevents
the second from forming. The more massive the first star to form, the bigger the effect. Figure~\ref{f:strominit}
shows the Str{\"o}mgren radius as a function of mass 
just after the star has formed. Two metallicities are shown: $Z$ = 0 and $Z$ = $1/50~$Z$_\odot$, while the metallicity in our model lies between these two values. 

While a 25~M$_\odot$ star has a Str{\"o}mgren sphere that covers half the scale radius, lower mass stars have a much smaller effect. Figure~\ref{f:m25spline}
shows a sharp fall in photon flux for stars below 25~M$_\odot$. At  $Z$ = $10^{-2}$~Z$_\odot$, the radiation from a 12~M$_\odot$ star over its entire 20~Myr lifetime is 
the same as a 25~M$_\odot$ star emits in 200~kyr. The pre-ionization phase for such a star can be neglected for simplicity as it will only affect a few cells. 

\section{Results and Discussion}

\begin{figure}[htb]
\begin{center}
\includegraphics[width=.45\textwidth]{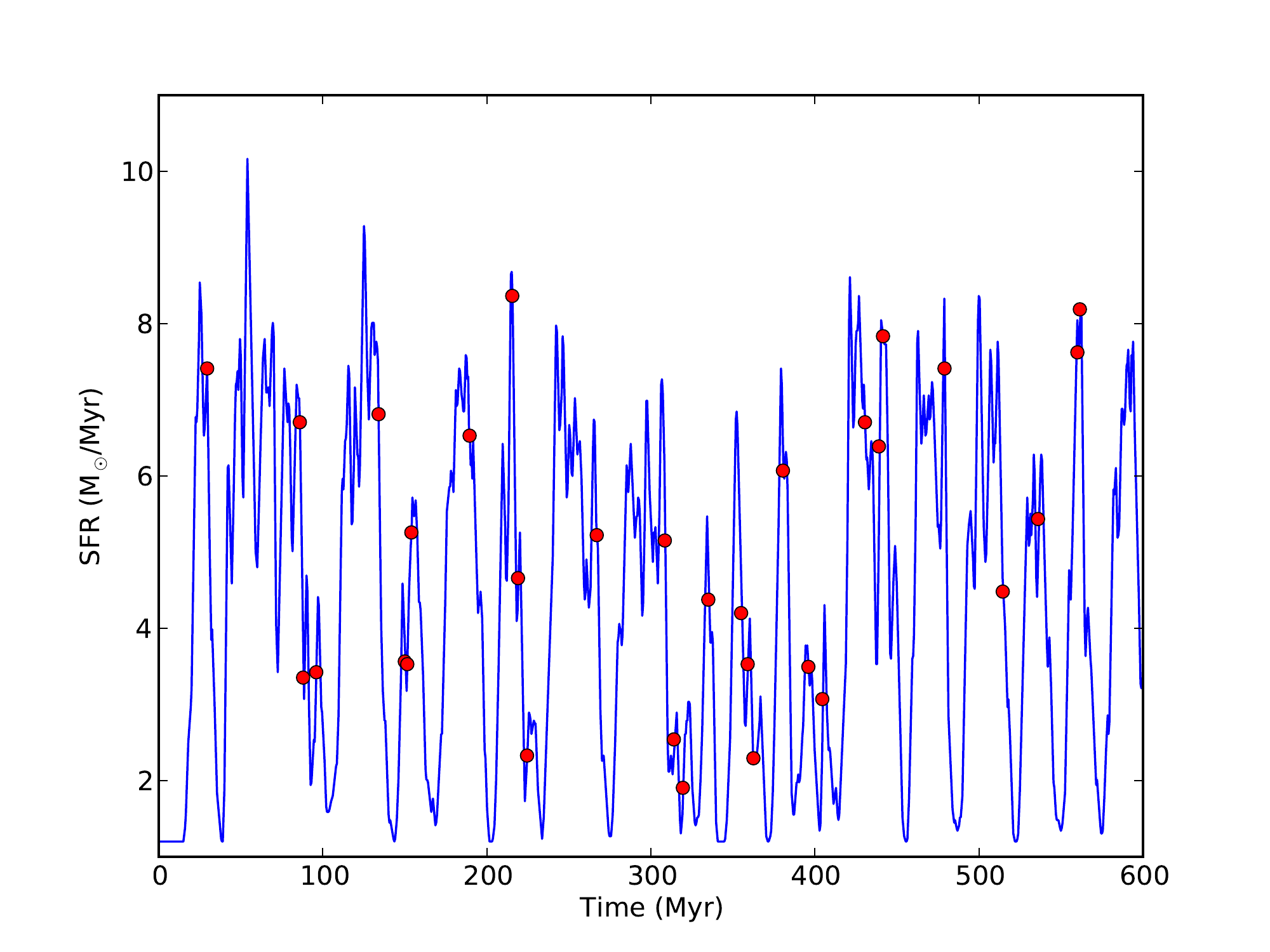}
\caption{Total star formation as a function of time, with the red circles corresponding to supernovae. The star formation is stochastic, leading to the plot being very noisy.}
\label{f:SFRt600}
\end{center}
\end{figure}

Figure~\ref{f:SFRt600} shows how the total star formation rate varies with time. The overall average is $\sim5$~M$_\odot$~Myr$^{-1}$, but with a 
large amount of noise due to the stochasticity of star formation. The effective Type II SN rate (including only the SN 
close enough to the center to have a 
significant impact on the system) is 0.05~SN~Myr$^{-1}$, or 1 SN/20~Myr.
The effective Type Ia SN rate is set to be 1 SN/20~Myr based on \citet{jiminez14}, such that after 100~Myr Type Ia SNe are as common as 
Type II.

\subsection{Description of a Typical Simulation Run}

\begin{figure*}[htb]
\begin{center}
\includegraphics[width=.9\textwidth]{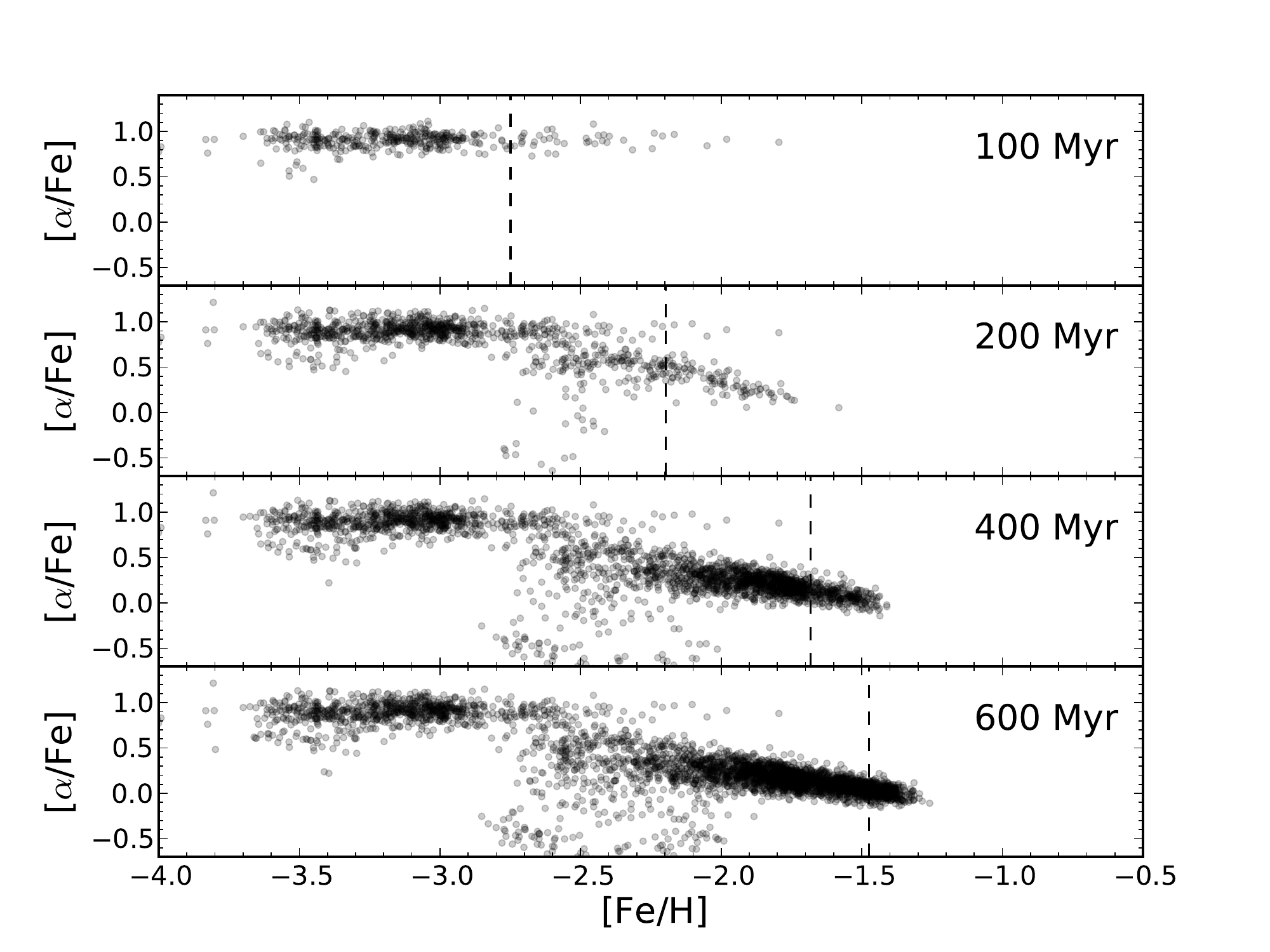}
\caption{Plots of [$\alpha$/Fe] vs [Fe/H] for stars after 100 (top), 200, 400 and 600~Myr. The dashed vertical lines in each panel are to the right of 90\% of the points and
can be considered a typical value of [Fe/H] at which stars form at that time.}
\label{f:afe}
\end{center}
\end{figure*}

In this section, we describe a sample simulation run, the results of which are shown in Figure~\ref{f:afe}. For the first 6~Myr, 
corresponding to the lifetime of the initial 25~M$_\odot$ star, [Fe/H] $\approx-4$ 
everywhere in the gas. As discussed in Section 3, we do not allow low-mass stars to form at this metallicity. 
At $t~=~6$~Myr, the 25~M$_\odot$ star explodes and
begins to enrich the gas. The first low-mass stars form at around 17~Myr and have a wide range of [Fe/H] from $-3.8$ to $-2$, while [$\alpha$/Fe] shows very little scatter and is
$\approx$~0.85. This is 
higher than the expected long-term average [$\alpha$/Fe] from Type II SNe, because a 25~M$_\odot$ star is more
massive than 80\% of Type II SN-forming stars and more massive stars have higher $\alpha$ yields relative to iron.
In this run, the first star with greater than 8~M$_\odot$ forms at 32~Myr, which is slightly higher than the median time of 25~Myr. This star has
 a mass of 9.7~M$_\odot$, corresponding to a lifetime of 29~Myr. At 38~Myr, a 9.0~M$_\odot$ star forms, which will explode just after the gas has recovered from the previous SN, and this is followed by a 10~M$_\odot$ star forming at 46~Myr. The combined effect of this is a period of low star formation from 60-100~Myr, although a few stars do form at a lower [$\alpha$/Fe] $\sim$ 0.5 on the side opposite to the later SNe. 
Note that in Fig.~\ref{f:afe} these stars are at 
low [Fe/H], as the gas with higher [Fe/H] has not yet recovered from the multiple SNe that enriched it. 

Between 100 and 200~Myr, enrichment from Type Ia SNe results in a decline in [$\alpha$/Fe] with increasing [Fe/H]. A few stars form with 
[$\alpha$/Fe] $<$ 0 in regions that are well away from the locations of all previous Type II SNe, but close to one or more Type Ia SNe. 
By 200~Myr, stars with [Fe/H] $>-2$ are becoming common and the mean [Fe/H] is $-2.9$, increasing to -$2.3$ at 400~Myr. We terminate star formation at
600~Myr as it is likely that star formation will be switched off by closely spaced SNe or the epoch of reionization at some time before this. 
The simulation ends with 2950~M$_\odot$ of stars
having formed, and there have been 29 Type II and 25 Type Ia SNe close enough to the center to significantly disrupt and enrich the gas.

The above description is an example only, and given that we assume star formation will occur stochastically, with variations in the number, mass and location of stars,
different runs of the simulation can give quite different results.
The total star formation and therefore number of SNe over the 600~Myr does not vary much between runs, meaning that the stars in the clump at 
[Fe/H] $\sim-1.5$
is a typical feature, although they can have slightly higher or lower [$\alpha$/Fe] due to variation in the relative number of Type Ia and Type II SNe. The mean [Fe/H] = $-2.10\pm0.09$ and the mean [$\alpha$/Fe] = 0.28$\pm$0.03. Note that these errors are simply the statistical errors based on a number 
of runs and do not take 
into account uncertainties in our star formation rate, 
IMF and Type Ia SN frequency. 
The large scatter in [$\alpha$/Fe] at [Fe/H] $\approx-2.5$ is a result of stars forming in regions that have been affected by very few SNe. At 
lower [Fe/H] there is little scatter because all stars have been affected by only a single SN, while at higher [Fe/H] there have been enough 
SNe that significant variations from the mean are uncommon.

\begin{figure*}[htb]
\begin{center}
\includegraphics[width=.9\textwidth]{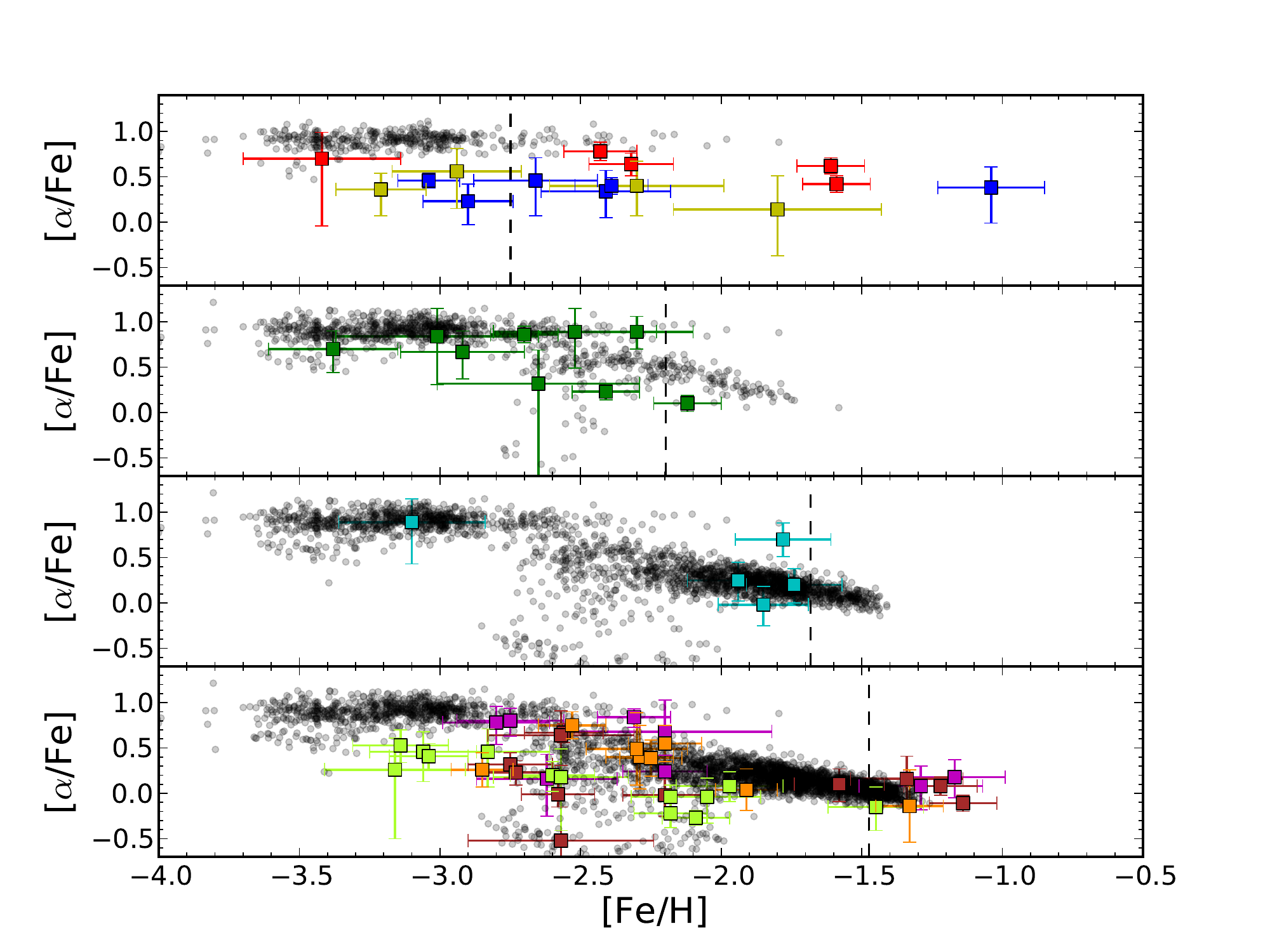}
\caption{Data points from \citet{vargas13} and \citet{kirby13} plotted on the stars produced by a simulation run. The top panel compares Segue I (red), Ursa Major II (blue) and Leo IV (yellow), the three systems that do not show evidence of Type Ia enrichment, plotted on our 100~Myr results. The second panel compares Coma Berenices (green) to our 200~Myr results. The third panel compares Leo T to our 400~Myr results. The bottom panel shows Canes Venatici II (magenta), Ursa Major I (brown), Hercules (light green) and Segue 2 (orange) compared to our 600~Myr results.}
\label{f:vargasmetals}
\end{center}
\end{figure*}

\section{Comparison to Observed Systems}

We now compare the results of our simulations to the UFDs discovered by \citet{simon07} and \citet{simon11}, which were studied in [$\alpha$/Fe] by \citet{vargas13}, as well as to the mass-metallicity relation \citep{kirby13b}. The Vargas UFDs
have masses ranging from 
$6\times 10^5$~M$_\odot$ to $1.2\times 10^7$~M$_\odot$ within the half-light radius \citep{wolf10}. The total halo masses are uncertain, but may be significantly greater than this, such that the virial halo masses are likely to range from being similar to the $10^7$~M$_\odot$ in our model to being many times greater. In Figure~\ref{f:vargasmetals}, we plot the
[Fe/H] versus [$\alpha$/Fe] data from the \citet{vargas13} study on the results of one of our simulation runs and determine whether our model can explain these systems. 
\citet{vargas13} note that the galaxies show evidence of old stellar populations without intermediate-age stars, indicative of a single burst of star formation lasting
less than 2~Gyr.

\citet{vargas13} give abundance ratios for 61 red giant branch stars in 8 UFDs, the largest sample of alpha abundances in galaxies this faint. 
They found that six of the eight show on average reduced [$\alpha$/Fe] at higher [Fe/H], meaning that stars have likely formed for longer than 100~Myr. 
Figure~\ref{f:vargasmetals} shows [$\alpha$/Fe] versus [Fe/H] for the complete Vargas sample, as well as Segue 2 from \citet{kirby13} with our model prediction superimposed. In this section we discuss whether our simulations can explain these systems. The definition of [$\alpha$/Fe] for the observed values 
uses the same four elements as used to define $\alpha$ in our models, although our method has the effect of giving a greater weight 
to [Si/Fe] and [Mg/Fe], 
which are the more abundant elements, and a very low weight to [Ti/Fe]. Titanium is underproduced in the \citep{woosley95} 
nucleosynthetic models with respect to the solar values, while the Ca, Si and Mg yields are consistent with the solar values and each other. 

\subsection{Systems Showing no Evidence of Type Ia Supernovae}

Segue 1 (plotted in red in Figure~\ref{f:vargasmetals}) has a
mass of 5.8$^{+8.2}_{-3.1}\times 10^5$~M$_\odot$ within the half-light radius.
Only five stars were observed by \citet{simon11}, with 
four having [$\alpha$/Fe] 
between 0.6 and 0.9 and the other 0.4. \citet{frebel14} observed two more stars in Segue 1 with 
[Fe/H] $< -3.5$  and suggested that it was the least evolved known galaxy.
The two stars with [Fe/H] $\approx$ 0.5 are unusual, being two of only
three stars in the Vargas sample at [Fe/H] $<$ -2 with [$\alpha$/Fe] $>$ 0.4. Our models show very few stars with [Fe/H] $> -2$ forming in the first 
100~Myr of the evolution of the system.

\begin{figure}[htb]
\begin{center}
\includegraphics[width=.45\textwidth]{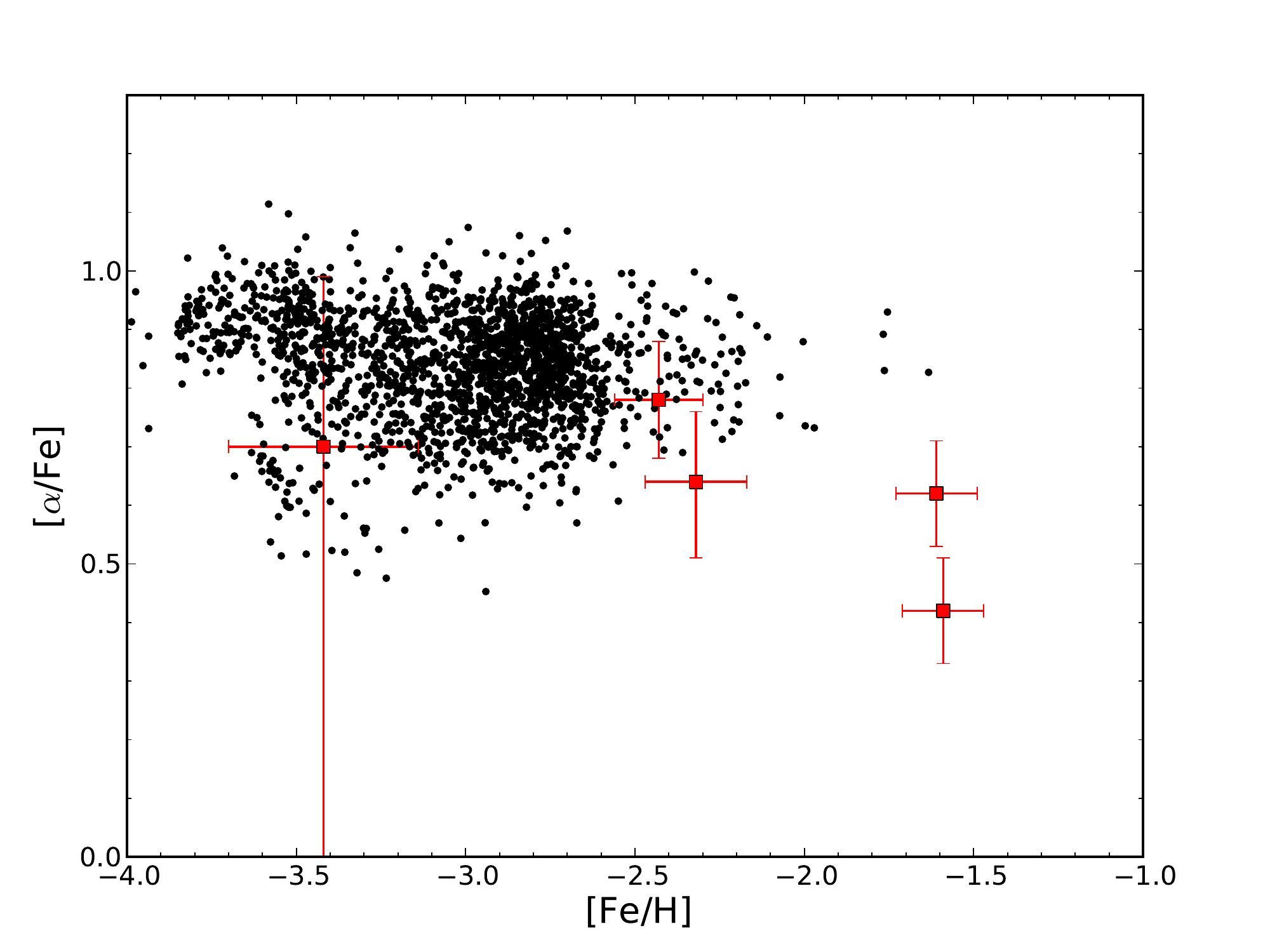}
\caption{Observations of Segue 1 from \citet{vargas13} (red points) compared to a 200 Myr run of our model where star formation is doubled and Type Ia supernovae are not allowed.}
\label{f:200myrresults}
\end{center}
\end{figure}

Ursa Major II (dark blue in Figure~\ref{f:vargasmetals}), which has a half-light radius mass of 7.9$^{+5.6}_{-3.1}\times 10^6$~M$_\odot$ \citep{wolf10}, also shows no evidence of a decline in [$\alpha$/Fe].
Given sufficient Type II SNe and the absence of Type Ia SNe, [$\alpha$/Fe] will eventually settle around 0.35 \citep{frebel12}. 
The stars in Ursa Major II are clustered around [$\alpha$/Fe] = 0.4, with all but one of the stars having [Fe/H] $<-2.5$. The exception is a star with [Fe/H] = $-1.1$ and 
[$\alpha$/Fe] = 0.4; however this may be a foreground star rather than a member of the system \citep{frebel10}.

Leo IV (yellow in Figure~\ref{f:vargasmetals}) has a half-light radius mass of 1.2$^{+3.5}_{-0.9}\times 10^6$~M$_\odot$ \citep{wolf10} and contains only four observed stars, all with large 
uncertainties in their abundance ratios. More stellar abundances are required to determine whether [$\alpha$/Fe] decreases with increasing [Fe/H] in 
this system.Leo IV has a luminosity of 8700~L$_\odot$, suggesting a factor of 40 more star formation than in our models for a typical mass-to-light ratio and a Kroupa 
IMF \citep[e.g][]{martin08} if it formed stars for $<100$~Myr. The system may be a closer match to our models if the period of star 
formation was longer. \citet{brown12} gives the constraint that the spread of stellar ages is less than 2~Gyr. 

The stars from these three systems are plotted on an example run of our simulation after 100~Myr in the top panel of Figure~\ref{f:vargasmetals}.  Segue 1, with its low luminosity and half-light 
radius mass, is the best fit to our models, however two
stars do not appear to fit with star formation times of $<$200 Myr. We note that there are large uncertainties associated with the star formation rate, the Type Ia SN rate, and the time of the first Type Ia SN. We therefore test increasing
the star formation rate by a factor of two and evolving for 200~Myr without allowing Type Ia SNe. The results are shown in Figure~\ref{f:200myrresults}. 
Increasing the undisturbed star formation rate 
has only a small effect on the total enrichment, because it leads to SNe occurring more quickly after the gas recovers, 
which leads to a period of low 
or no star formation. A larger halo mass would be required to maintain a higher average star formation rate. Frebel et al. (2014) calculated the initial stellar mass for Segue 1 to be $\sim1500$~M~$_\odot$, which suggests a star formation rate a factor of two to three higher than for our models if the length of star formation was 100-150~Myr. However, this is not a large enough factor to explain the 
outliers at [Fe/H]~$\approx-1.5$.


\begin{figure*}[htb]
\begin{center}
\includegraphics[width=.9\textwidth]{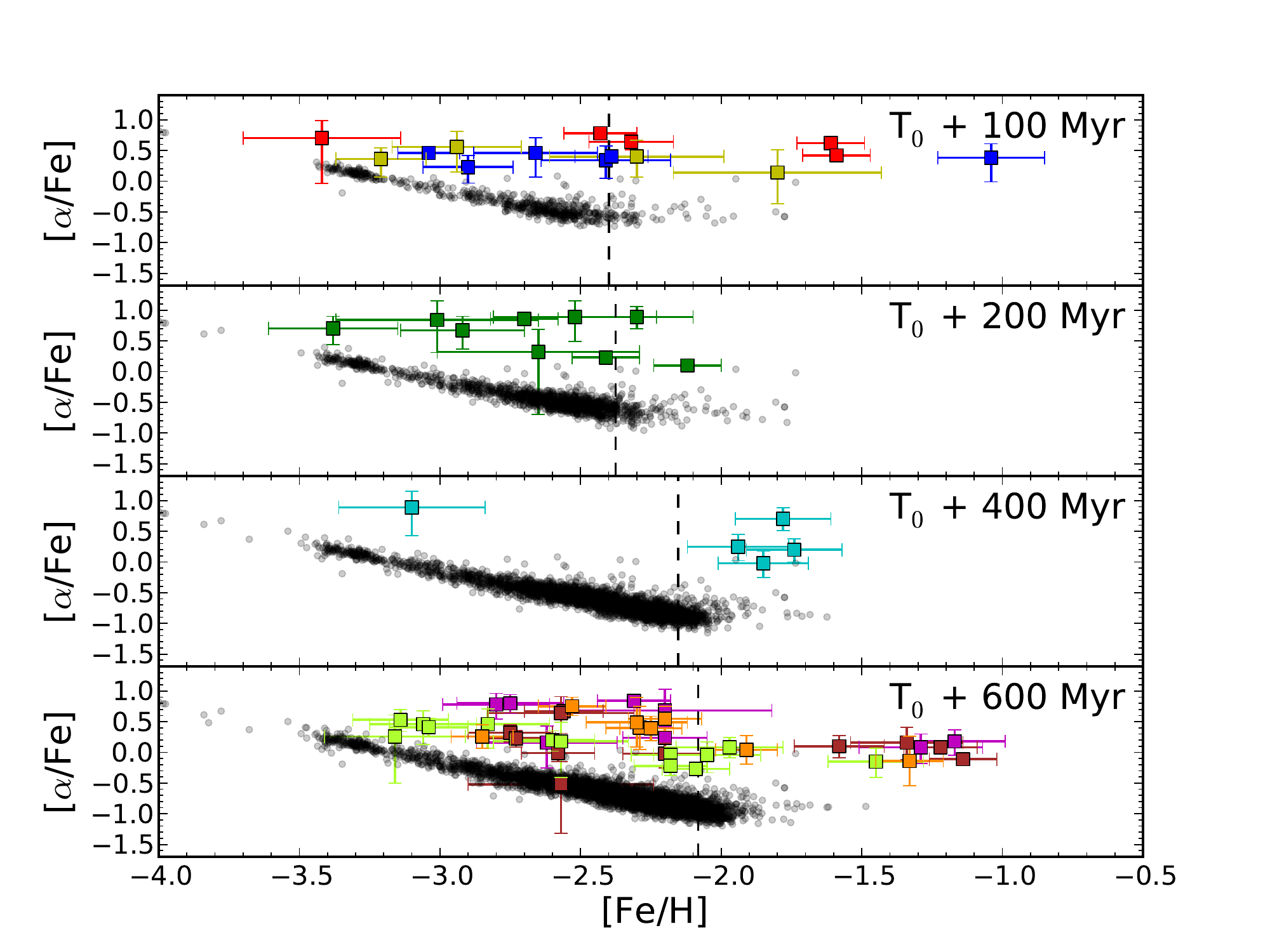}
\caption{Data points from \citet{vargas13} and \citet{kirby13} plotted on the stars produced by a simulation run with Type Ia-only enrichment. $T_0 = $100~Myr is 
the delay time from the formation of the first star to the time of the first Type Ia supernova. 
The observations are as for Figure~\ref{f:vargasmetals}: the top panel compares Segue I (red), Ursa Major II (blue) and Leo IV (yellow), the three systems that do not show evidence of Type Ia enrichment, plotted on our 100~Myr results. The second panel compares Coma Berenices (green) to our 200~Myr results. The third panel compares Leo T to our 400~Myr results. The bottom panel shows Canes Venatici II (magenta), Ursa Major I (brown), Hercules (light green) and Segue 2 (orange) compared to our 600~Myr results.}
\label{f:typeiaplot}
\end{center}
\end{figure*}

\subsection{Systems showing evidence of Type Ia supernovae}

The remaining systems show evidence of Type Ia SNe, containing stars with solar or sub-solar [$\alpha$/Fe]. These galaxies have masses in the range
$6\times10^5~-~1.2\times10^7$~M$_\odot$ within the half-light radius \citep{wolf10}. All show lower [$\alpha$/Fe] for higher [Fe/H], which is 
interpreted as the effect of Type Ia SNe, which yield much more iron relative to $\alpha$ elements compared to Type II SNe. The decline 
in [$\alpha$/Fe] may commence before the first Type Ia SNe due to Type II SNe from lower mass (8-15~M$_\odot$) stars. Averaged over a Kroupa IMF,
the yield from Type II SNe [$\alpha$/Fe] $\approx0.35$, significantly lower than for many stars in the \citet{vargas13} sample.   

In the \citet{vargas13} sample, nine stars are observed in Coma Berenices, of which seven show [Fe/H] $<-2.5$, with six of these clustering around [$\alpha$/Fe] $\approx 0.8$.
The other two stars show [Fe/H] $<-2$ and [$\alpha$/Fe] $\approx$ 0.2. This could be explained by a few high-mass stars affecting the system in the first 
100~Myr, followed by a small number of Type Ia SNe, after which star formation quickly ends. The observed stars in Coma Berenices 
(dark green in 
\ref{f:vargasmetals})
have a mean [Fe/H] of $-2.7$ and 
we therefore compare it to our 200~Myr plot in Figure~\ref{f:vargasmetals}, which has mean [Fe/H] = $-2.9$. The metallicity observations fit well, 
suggesting that Coma Berenices may have formed stars for $\sim$200~Myr. The luminosity of 3700~L$_\odot$ suggests $\sim5-10$ times more 
star formation than for our model if this is the case. The mass within the half-light radius is 
$2.0^{+0.9}_{-0.6}\times 10^6$~M$_\odot$ \citep{wolf10} and it does not show 
evidence of tidal disruption \citep{munoz10}.

Leo T is shown in the third panel of Figure~\ref{f:vargasmetals}. There are only five stars, four of them with $-2.0~<$~[Fe/H]~$<~-1.5$,
consistent with a system that formed stars for $\sim$400~Myr. The star with [Fe/H] = $-1.8$ and [$\alpha$/Fe] = 0.7 is rare in our models and requires enrichment by multiple Type II SN from high mass stars without significant enrichment from Type Ia SN. Leo T is the most luminous system in the Vargas sample and the only one that shows evidence of star formation at late times.

The rest of the Vargas systems are shown in the bottom panel of Figure~\ref{f:vargasmetals} along with Segue 2 from \citet{kirby13}. 
All contain stars with [Fe/H]~$>~-1.5$ and some show stars with [Fe/H]~$>-1.1$. 

These systems have star formation histories of less than 2~Gyr \citep{vargas13}. Ursa Major I and Hercules, with their luminosities of $>10^4$~L$_\odot$ and their half-light halo 
masses $\sim 10^7$~M$_\odot$, are likely significantly more massive than our modeled systems. Canes Venatici II, with a half-light radius mass of  $1.4^{+1.0}_{-0.6}\times 10^6$~M$_\odot$ and a luminosity of  $7.9^{+4.4}_{-3.0}\times 10^3$~L$_\odot$ \citep{wolf10},
is a closer fit, although the implied $>10^4$~M$_\odot$ of stars is still more than in our model.


The remaining systems do not contain stars with [Fe/H] $<-3$, but other features match well with our model, with nearly all 
observed stars lying close to regions with a high density of modeled stars. 
The low-metallicity stars in Hercules do not fall on the clump of modeled stars, but can be explained by the effects of one or more early 
Type II SN from lower mass stars, which would have the effect of reducing [$\alpha$/Fe]. The large scatter in [$\alpha$/Fe] near [Fe/H] = -2.5 
in the model is seen in Ursa Major I and Canes Venatici II, although not in Hercules or Segue 2. In our simulation the scatter results from 
the range of $\alpha$ enrichment from different-mass Type II 
SNe, as well as the commencement of Type Ia SNe.

Within each galaxy there is much less scatter in [$\alpha$/Fe] than in [Fe/H], especially at low [Fe/H], which is expected for the reasons outlined in \citet{frebel12}. 
Ursa Major I contains a star at [Fe/H] = $-2.6$ with approximately solar [$\alpha$/Fe], which could be indicative of a low star formation rate resulting in lower enrichment
for the first 100~Myr, but could also be explained simply by a case where fewer SNe occur in the first 100~Myr due to the stochastic nature of 
stellar masses.

Segue 2 is the least massive known galaxy, with $<1.5\times 10^{5}$~M$_\odot$ within the half-light radius, a luminosity of 900~L$_\odot$ 
and a stellar mass of 1000~M$_\odot$. This is much less massive than our modeled systems. \citet{belokurov09} give a total halo mass of 
$5.5\times10^5$~M$_\odot$; however, it should be noted that this is highly model-dependent and the true mass could be higher.
\citet{kirby13} discuss two possibilities, the first being that the galaxy formed at a mass similar to the 
observed mass, and the second being that the system formed with a mass $\sim 10^9$~M$_\odot$, but that 99.7\% of the stars have been stripped. 
The second scenario requires a highly eccentric orbit \citep[50:1][]{penarrubia08}
This would be a highly unusual object and \citet{kirby13} note that it would be the first known galaxy to 
have lost its dark matter halo without being completely disrupted. While Segue 2 is the only significant outlier from the stellar 
mass-metallicity relation \citep{kirby13b}, which supports the tidal stripping scenario, \citet{chen14} suggest 
that the stellar mass-metallicity relation becomes flat or even changes direction below $M{\rm_{vir}} = 10^8$~M$_\odot$, because lower mass halos 
are less efficient in accreting metal-poor gas. The current observational data cannot distinguish between the possible origins of Segue 2. 

The observed stars from Segue 2 are plotted in orange on the bottom panel in Figure~\ref{f:vargasmetals}. Six of the eight stars show [$\alpha$/Fe] $>$ 0.3 and [Fe/H] $<-2.2$.
One of the remaining stars is at [Fe/H] = $-1.9$ and solar [$\alpha$/Fe], while the other is at [Fe/H] = $-1.3$ and sub-solar [$\alpha$/Fe]. 

We also tested a scenario without enrichment from Type II SNe, where the Type Ia SN rate was increased such that the total 
SN rate remained the same. The implicit assumption made here is that intermediate-mass (3-8~M$_\odot$) stars can form at our starting metallicity of [Fe/H] = $-4$. The results are shown in Figure~\ref{f:typeiaplot} and show that the decline in [$\alpha$/Fe] is too rapid to explain UFD 
observations. Within 200~Myr, most stars formed in our simulation have lower [$\alpha$/Fe] than any observed stars. We therefore conclude that enrichment
from Type II SNe is required to explain the metallicity of stars in the UFDs observed to date. 

However, given the results of Paper I, 
a $10^{6.5}$~M$_\odot$ galaxy would likely survive a single Type Ia SN, which does not have the pre-ionization phase associated with 
Type II SNe. The star formation rate of M65 in its undisturbed state is 5.7~M$_\odot$Myr$^{-1}$, which corresponds to an average of 10 stars per Myr. In the Kroupa IMF, 1 in 270 stars is greater than 8 solar masses, so the average SN rate if the gas is not disturbed is 1 per 27 Myr; however, the stochastic nature of star 
formation means that in 15\% of cases no stars with greater than 8~M$_\odot$ will form within 50~Myr and in 
2\% of cases none will form within the first 100~Myr. Such systems could therefore experience one or more prompt Type Ia SNe
\citep{mannucci06} before any Type II SN.
 
Only a few hundred stars would form in such a system, and they may be beyond the reach of current telescopes. Future detections of stars in such 
systems are likely to show very low [$\alpha$/Fe] relative to [Fe/H] because the surviving systems will be those that lacked the 
$\alpha$-enriching massive stars. Systems that form with a high enough metallicity for immediate low-mass star formation may show a few very low [Fe/H] stars with high [$\alpha$/Fe]. Such stars would form in gas enriched only by Population III stars.

\subsection{Scaling Relations}

Dwarf galaxies observed to date obey a tight relationship between stellar mass and metallicity, 

\begin{equation}
\rm{[Fe/H]} = -1.69 + 0.30\log{(\frac{M_*}{10^6~\rm{M}_\odot})}
\end{equation} 

with an rms of 0.16 \citep{kirby13b}.

After 400~Myr in our model, the stellar mass is $\approx2000$~M$_\odot$ and [Fe/H] = $-2.3\pm0.1$. Assuming that 
due to stellar evolution the stellar mass observed today would be half of its original value (as calculated for Segue 1 by \citet{frebel14}), the 
mass-metallicity relation gives [Fe/H] = $-2.59\pm0.16$. The discrepancy is larger at 600~Myr, where [Fe/H] = $-2.1\pm0.1$ for our model compared to 
[Fe/H] = $-2.54\pm0.16$ from the mass-metallicity relation. 

However, our models do fit well with the simulations of \citet{chen14}, who suggest that there is a change of slope in a number of scaling 
relations at $M_{\rm{vir}} \sim 10^8$~M$_\odot$ resulting from the inefficiency of low-mass halos in accreting metal-poor gas. 
The mean stellar metallicity of $-2.1$ at 600~Myr in our model is slightly lower than they find for a 
$M_{\rm{vir}} = 10^7$~M$_\odot$ halo, although is still within their range of uncertainty. They find that the total stellar mass is equal to:

\begin{equation}
\rm{log}M_* = 3.5 + 1.3\log{(M_{\rm{vir}}/10^7~\rm{M}_\odot)}
\end{equation}

This gives M$_*$ = 3200~M$_\odot$, which is similar to the stellar mass of 3000~M$_\odot$ from our models.


\section{Summary}

We have simulated star formation and chemical enrichment in systems with dark matter halo masses of $10^7$~M$_\odot$, which is a mass similar to many of the recently discovered UFDs.
We use the single SN simulations of Paper I as a starting point and evolve the system for 600~Myr. Using a simple model of star formation,
we find the following:
\begin{enumerate}
\item A single star with a mass of $25$~M$_\odot$ is sufficient to permanently stop star formation in a system with a mass of $10^{6.5}$~M$_\odot$, even if the star is as much as 0.5 scale radii 
(50~pc) from the center. 
\item In systems with $10^7$~M$_\odot$, massive stars farther than 0.5 scale radii (75~pc) from the center have little impact on the system as most of the energy and metals
escape.
\item Systems with $10^7$~M$_\odot$ recover from the first SN explosion and form low-mass stars for more than 100~Myr.
\item The assumption that star formation occurs at the same rate as in the Carina dwarf (scaled for the size of our modeled systems) predicts a turnover 
of [$\alpha$/Fe] at [Fe/H] $\sim-2.5$.
\item Our models predict $>$1~dex of scatter in [$\alpha$/Fe] close to the turnover, but that away from this value of [Fe/H] the scatter should be 
$<0.5$~dex.
\item In terms of stellar and halo mass, Segue 1 is the closest observed match to our modeled systems. The stellar mass of $\sim1500$~M$_\odot$ calculated by \citet{frebel14}, along with the lack of evidence of enrichment from Type Ia SNe, suggests a star formation rate a factor of three to four times higher than 
for our models, while the halo mass estimate from \citet{simon11} suggests the system is $\sim$3 times more massive than our M70 model. 
With better instruments, we expect the detection of many systems similar to Segue 1, which is currently the least luminous known galaxy. 
This paper provides support to the suggestion of \citet{bovill09} that at least some such systems are fossil galaxies that formed in the early universe with halo masses $<10^8$~M$_\odot$, then experienced only a single burst of star formation.
\item We find that systems with $M_{vir} = 10^{6.5}$~M$_\odot$ and lower cannot survive the feedback from their star formation. This 
suggests that if Segue 2 has a virial mass of $<10^6$~M$_\odot$ as suggested by \citet{kirby13}, it could not have formed with its present halo mass, 
but rather must have experienced at least some tidal stripping. However, if its true virial mass is $\gtrsim10^{6.5}$~M$_\odot$, it is consistent with 
the stellar mass-metallicity relation of \citet{chen14} and may therefore be an intrinsically very low mass system.
\item All observed stars in the rest of the Vargas UFDs have metallicities that are consistent with our model, 
although our model predicts stars with 
lower metallicities than have been observed. This discrepancy may result from the starting metallicity of [Fe/H] = $-4$, or the metallicity floor for 
low-mass star formation.

\end{enumerate} 

We have assumed our systems are formed and evolve in isolation and have not addressed the complex issues of accretion, ram pressure stripping, 
dark matter halo stripping, 
and other effects that may affect the evolution of the galaxy. 
While our modeled systems are in isolation, the UFDs compared to in this work have been affected by the M31 and Milky Way environment,
 as only the closest systems are 
bright enough to be observed. In future work we will seek 
to use our simulations to model very metal-poor damped Ly-$\alpha$ systems at $z~=~2$ \citep{cooke10,cooke11,cooke12}.

\acknowledgments DW is funded by an Australian Postgraduate Award. JBH is funded by an ARC Laureate Fellowship.

\bibliographystyle{apj}
\bibliography{refs}

\begin{thebibliography}{66}
\expandafter\ifx\csname natexlab\endcsname\relax\def\natexlab#1{#1}\fi

\bibitem[{{Abel} {et~al.}(2002){Abel}, {Bryan}, \& {Norman}}]{abel02}
{Abel}, T., {Bryan}, G.~L., \& {Norman}, M.~L. 2002, Science, 295, 93

\bibitem[{{Argast} {et~al.}(2000){Argast}, {Samland}, {Gerhard}, \&
  {Thielemann}}]{argast00}
{Argast}, D., {Samland}, M., {Gerhard}, O.~E., \& {Thielemann}, F.-K. 2000,
  \aap, 356, 873

\bibitem[{{Barkana} \& {Loeb}(1999)}]{barkana99}
{Barkana}, R., \& {Loeb}, A. 1999, \apj, 523, 54

\bibitem[{{Belokurov} {et~al.}(2009){Belokurov}, {Walker}, {Evans}, {Gilmore},
  {Irwin}, {Mateo}, {Mayer}, {Olszewski}, {Bechtold}, \&
  {Pickering}}]{belokurov09}
{Belokurov}, V., {et~al.} 2009, \mnras, 397, 1748

\bibitem[{{Bigiel} {et~al.}(2008){Bigiel}, {Leroy}, {Walter}, {Brinks}, {de
  Blok}, {Madore}, \& {Thornley}}]{bigiel08}
{Bigiel}, F., {Leroy}, A., {Walter}, F., {Brinks}, E., {de Blok}, W.~J.~G.,
  {Madore}, B., \& {Thornley}, M.~D. 2008, \aj, 136, 2846

\bibitem[{{Bland-Hawthorn} {et~al.}(2010){Bland-Hawthorn}, {Karlsson},
  {Sharma}, {Krumholz}, \& {Silk}}]{bland10}
{Bland-Hawthorn}, J., {Karlsson}, T., {Sharma}, S., {Krumholz}, M., \& {Silk},
  J. 2010, \apj, 721, 582

\bibitem[{{Bland-Hawthorn} {et~al.}(2011){Bland-Hawthorn}, {Sutherland}, \&
  {Karlsson}}]{bland11}
{Bland-Hawthorn}, J., {Sutherland}, R., \& {Karlsson}, T. 2011, in EAS
  Publications Series, Vol.~48, EAS Publications Series, ed. M.~{Koleva},
  P.~{Prugniel}, \& I.~{Vauglin}, 397--404

\bibitem[{{Bland-Hawthorn} {et~al.}(2014){Bland-Hawthorn}, {Sutherland}, \&
  {Webster}}]{bland14}
{Bland-Hawthorn}, J., {Sutherland}, R., \& {Webster}, D. 2014, submitted

\bibitem[{{Bovill} \& {Ricotti}(2009)}]{bovill09}
{Bovill}, M.~S., \& {Ricotti}, M. 2009, \apj, 693, 1859

\bibitem[{{Bovill} \& {Ricotti}(2011)}]{bovill11a}
---. 2011, \apj, 741, 17

\bibitem[{{Bromm} \& {Larson}(2004)}]{bromm04}
{Bromm}, V., \& {Larson}, R.~B. 2004, \araa, 42, 79

\bibitem[{{Bromm} \& {Yoshida}(2011)}]{bromm11}
{Bromm}, V., \& {Yoshida}, N. 2011, \araa, 49, 373

\bibitem[{{Bromm} {et~al.}(2003){Bromm}, {Yoshida}, \& {Hernquist}}]{bromm03}
{Bromm}, V., {Yoshida}, N., \& {Hernquist}, L. 2003, \apjl, 596, L135

\bibitem[{{Brown} {et~al.}(2012){Brown}, {Tumlinson}, {Geha}, {Kirby},
  {VandenBerg}, {Mu{\~n}oz}, {Kalirai}, {Simon}, {Avila}, {Guhathakurta},
  {Renzini}, \& {Ferguson}}]{brown12}
{Brown}, T.~M., {et~al.} 2012, \apjl, 753, L21

\bibitem[{{Castelli} \& {Kurucz}(2004)}]{castelli04}
{Castelli}, F., \& {Kurucz}, R.~L. 2004, ArXiv Astrophysics e-prints

\bibitem[{{Chen} {et~al.}(2014){Chen}, {Wise}, {Norman}, {Xu}, \&
  {O'Shea}}]{chen14}
{Chen}, P., {Wise}, J.~H., {Norman}, M.~L., {Xu}, H., \& {O'Shea}, B.~W. 2014,
  ArXiv e-prints

\bibitem[{{Cooke} {et~al.}(2012){Cooke}, {Pettini}, \& {Murphy}}]{cooke12}
{Cooke}, R., {Pettini}, M., \& {Murphy}, M.~T. 2012, \mnras, 425, 347

\bibitem[{{Cooke} {et~al.}(2010){Cooke}, {Pettini}, {Steidel}, {King}, {Rudie},
  \& {Rakic}}]{cooke10}
{Cooke}, R., {Pettini}, M., {Steidel}, C.~C., {King}, L.~J., {Rudie}, G.~C., \&
  {Rakic}, O. 2010, \mnras, 409, 679

\bibitem[{{Cooke} {et~al.}(2011){Cooke}, {Pettini}, {Steidel}, {Rudie}, \&
  {Nissen}}]{cooke11}
{Cooke}, R., {Pettini}, M., {Steidel}, C.~C., {Rudie}, G.~C., \& {Nissen},
  P.~E. 2011, \mnras, 417, 1534

\bibitem[{{Dopita} \& {Sutherland}(2003)}]{dopita03}
{Dopita}, M.~A., \& {Sutherland}, R.~S. 2003, {Astrophysics of the diffuse
  universe}

\bibitem[{{Ekstr{\"o}m} {et~al.}(2006){Ekstr{\"o}m}, {Meynet}, \&
  {Maeder}}]{ekstrom2006}
{Ekstr{\"o}m}, S., {Meynet}, G., \& {Maeder}, A. 2006, in Astronomical Society
  of the Pacific Conference Series, Vol. 353, Stellar Evolution at Low
  Metallicity: Mass Loss, Explosions, Cosmology, ed. H.~J.~G.~L.~M. {Lamers},
  N.~{Langer}, T.~{Nugis}, \& K.~{Annuk}, 141

\bibitem[{{Frebel} \& {Bromm}(2012)}]{frebel12}
{Frebel}, A., \& {Bromm}, V. 2012, \apj, 759, 115

\bibitem[{{Frebel} {et~al.}(2007){Frebel}, {Johnson}, \& {Bromm}}]{frebel07}
{Frebel}, A., {Johnson}, J.~L., \& {Bromm}, V. 2007, \mnras, 380, L40

\bibitem[{{Frebel} {et~al.}(2010){Frebel}, {Simon}, {Geha}, \&
  {Willman}}]{frebel10}
{Frebel}, A., {Simon}, J.~D., {Geha}, M., \& {Willman}, B. 2010, \apj, 708, 560

\bibitem[{{Frebel} {et~al.}(2014){Frebel}, {Simon}, \& {Kirby}}]{frebel14}
{Frebel}, A., {Simon}, J.~D., \& {Kirby}, E.~N. 2014, \apj, 786, 74

\bibitem[{{Fumagalli} {et~al.}(2011){Fumagalli}, {O'Meara}, \&
  {Prochaska}}]{fumagalli11}
{Fumagalli}, M., {O'Meara}, J.~M., \& {Prochaska}, J.~X. 2011, Science, 334,
  1245

\bibitem[{{Georgy} {et~al.}(2012){Georgy}, {Ekstr{\"o}m}, {Meynet}, {Massey},
  {Levesque}, {Hirschi}, {Eggenberger}, \& {Maeder}}]{georgy2012}
{Georgy}, C., {Ekstr{\"o}m}, S., {Meynet}, G., {Massey}, P., {Levesque}, E.~M.,
  {Hirschi}, R., {Eggenberger}, P., \& {Maeder}, A. 2012, \aap, 542, A29

\bibitem[{{Gnedin}(2000)}]{gnedin00}
{Gnedin}, N.~Y. 2000, \apj, 542, 535

\bibitem[{{Greif} {et~al.}(2010){Greif}, {Glover}, {Bromm}, \&
  {Klessen}}]{greif10}
{Greif}, T.~H., {Glover}, S.~C.~O., {Bromm}, V., \& {Klessen}, R.~S. 2010,
  \apj, 716, 510

\bibitem[{{Hopkins} {et~al.}(2013){Hopkins}, {Narayanan}, \&
  {Murray}}]{hopkins13}
{Hopkins}, P.~F., {Narayanan}, D., \& {Murray}, N. 2013, \mnras, 432, 2647

\bibitem[{{Iwamoto} {et~al.}(1999){Iwamoto}, {Brachwitz}, {Nomoto},
  {Kishimoto}, {Umeda}, {Hix}, \& {Thielemann}}]{iwamoto99}
{Iwamoto}, K., {Brachwitz}, F., {Nomoto}, K., {Kishimoto}, N., {Umeda}, H.,
  {Hix}, W.~R., \& {Thielemann}, F.-K. 1999, \apjs, 125, 439

\bibitem[{{Jimenez} {et~al.}(2014){Jimenez}, {Tissera}, \&
  {Matteucci}}]{jiminez14}
{Jimenez}, N., {Tissera}, P.~B., \& {Matteucci}, F. 2014, ArXiv e-prints

\bibitem[{{Johnson} {et~al.}(2008){Johnson}, {Greif}, \& {Bromm}}]{johnson08}
{Johnson}, J.~L., {Greif}, T.~H., \& {Bromm}, V. 2008, \mnras, 388, 26

\bibitem[{{Karlsson} {et~al.}(2012){Karlsson}, {Bland-Hawthorn}, {Freeman}, \&
  {Silk}}]{karlsson12}
{Karlsson}, T., {Bland-Hawthorn}, J., {Freeman}, K.~C., \& {Silk}, J. 2012,
  \apj, 759, 111

\bibitem[{{Karlsson} {et~al.}(2013){Karlsson}, {Bromm}, \&
  {Bland-Hawthorn}}]{karlsson13}
{Karlsson}, T., {Bromm}, V., \& {Bland-Hawthorn}, J. 2013, Reviews of Modern
  Physics, 85, 809

\bibitem[{{Kirby} {et~al.}(2013{\natexlab{a}}){Kirby}, {Boylan-Kolchin},
  {Cohen}, {Geha}, {Bullock}, \& {Kaplinghat}}]{kirby13}
{Kirby}, E.~N., {Boylan-Kolchin}, M., {Cohen}, J.~G., {Geha}, M., {Bullock},
  J.~S., \& {Kaplinghat}, M. 2013{\natexlab{a}}, \apj, 770, 16

\bibitem[{{Kirby} {et~al.}(2013{\natexlab{b}}){Kirby}, {Cohen}, {Guhathakurta},
  {Cheng}, {Bullock}, \& {Gallazzi}}]{kirby13b}
{Kirby}, E.~N., {Cohen}, J.~G., {Guhathakurta}, P., {Cheng}, L., {Bullock},
  J.~S., \& {Gallazzi}, A. 2013{\natexlab{b}}, \apj, 779, 102

\bibitem[{{Kroupa}(2001)}]{kroupa01}
{Kroupa}, P. 2001, \mnras, 322, 231

\bibitem[{{Kudritzki}(2002)}]{kudritzki2002}
{Kudritzki}, R.~P. 2002, \apj, 577, 389

\bibitem[{{Kudritzki}(2005)}]{kudritzki2005}
{Kudritzki}, R.-P. 2005, in Astronomical Society of the Pacific Conference
  Series, Vol. 332, The Fate of the Most Massive Stars, ed. R.~{Humphreys} \&
  K.~{Stanek}, 239

\bibitem[{{Mac Low} \& {Ferrara}(1999)}]{maclow99}
{Mac Low}, M.-M., \& {Ferrara}, A. 1999, \apj, 513, 142

\bibitem[{{Madau} {et~al.}(2001){Madau}, {Ferrara}, \& {Rees}}]{madau01}
{Madau}, P., {Ferrara}, A., \& {Rees}, M.~J. 2001, \apj, 555, 92

\bibitem[{{Mannucci} {et~al.}(2006){Mannucci}, {Della Valle}, \&
  {Panagia}}]{mannucci06}
{Mannucci}, F., {Della Valle}, M., \& {Panagia}, N. 2006, \mnras, 370, 773

\bibitem[{{Martin} {et~al.}(2008){Martin}, {de Jong}, \& {Rix}}]{martin08}
{Martin}, N.~F., {de Jong}, J.~T.~A., \& {Rix}, H.-W. 2008, \apj, 684, 1075

\bibitem[{{Mayer} {et~al.}(2001){Mayer}, {Governato}, {Colpi}, {Moore},
  {Quinn}, {Wadsley}, {Stadel}, \& {Lake}}]{mayer01}
{Mayer}, L., {Governato}, F., {Colpi}, M., {Moore}, B., {Quinn}, T., {Wadsley},
  J., {Stadel}, J., \& {Lake}, G. 2001, \apj, 559, 754

\bibitem[{{Meynet} \& {Maeder}(2000)}]{meynet2000}
{Meynet}, G., \& {Maeder}, A. 2000, \aap, 361, 101

\bibitem[{{Meynet} \& {Maeder}(2002)}]{meynet2002}
---. 2002, \aap, 390, 561

\bibitem[{{Mu{\~n}oz} {et~al.}(2010){Mu{\~n}oz}, {Geha}, \&
  {Willman}}]{munoz10}
{Mu{\~n}oz}, R.~R., {Geha}, M., \& {Willman}, B. 2010, \aj, 140, 138

\bibitem[{{Muratov} {et~al.}(2013){Muratov}, {Gnedin}, {Gnedin}, \&
  {Zemp}}]{muratov13}
{Muratov}, A.~L., {Gnedin}, O.~Y., {Gnedin}, N.~Y., \& {Zemp}, M. 2013, \apj,
  773, 19

\bibitem[{{Nomoto} {et~al.}(2006){Nomoto}, {Tominaga}, {Umeda}, {Kobayashi}, \&
  {Maeda}}]{nomoto06}
{Nomoto}, K., {Tominaga}, N., {Umeda}, H., {Kobayashi}, C., \& {Maeda}, K.
  2006, Nuclear Physics A, 777, 424

\bibitem[{{Okamoto} {et~al.}(2008){Okamoto}, {Gao}, \& {Theuns}}]{okamoto08}
{Okamoto}, T., {Gao}, L., \& {Theuns}, T. 2008, \mnras, 390, 920

\bibitem[{{Pe{\~n}arrubia} {et~al.}(2008){Pe{\~n}arrubia}, {Navarro}, \&
  {McConnachie}}]{penarrubia08}
{Pe{\~n}arrubia}, J., {Navarro}, J.~F., \& {McConnachie}, A.~W. 2008, \apj,
  673, 226

\bibitem[{{Rees}(1986)}]{rees86}
{Rees}, M.~J. 1986, \mnras, 218, 25P

\bibitem[{{Ricotti}(2009)}]{ricotti09}
{Ricotti}, M. 2009, \mnras, 392, L45

\bibitem[{{Schaerer}(2002)}]{schaerer02}
{Schaerer}, D. 2002, \aap, 382, 28

\bibitem[{{Simon} \& {Geha}(2007)}]{simon07}
{Simon}, J.~D., \& {Geha}, M. 2007, \apj, 670, 313

\bibitem[{{Simon} {et~al.}(2011){Simon}, {Geha}, {Minor}, {Martinez}, {Kirby},
  {Bullock}, {Kaplinghat}, {Strigari}, {Willman}, {Choi}, {Tollerud}, \&
  {Wolf}}]{simon11}
{Simon}, J.~D., {et~al.} 2011, \apj, 733, 46

\bibitem[{{Sutherland} \& {Dopita}(1993)}]{sutherland93}
{Sutherland}, R.~S., \& {Dopita}, M.~A. 1993, \apjs, 88, 253

\bibitem[{{Thornton} {et~al.}(1998){Thornton}, {Gaudlitz}, {Janka}, \&
  {Steinmetz}}]{thornton98}
{Thornton}, K., {Gaudlitz}, M., {Janka}, H.-T., \& {Steinmetz}, M. 1998, \apj,
  500, 95

\bibitem[{{Tinsley}(1979)}]{tinsley79}
{Tinsley}, B.~M. 1979, \apj, 229, 1046

\bibitem[{{Tolstoy} {et~al.}(2009){Tolstoy}, {Hill}, \& {Tosi}}]{tolstoy09}
{Tolstoy}, E., {Hill}, V., \& {Tosi}, M. 2009, \araa, 47, 371

\bibitem[{{Vargas} {et~al.}(2013){Vargas}, {Geha}, {Kirby}, \&
  {Simon}}]{vargas13}
{Vargas}, L.~C., {Geha}, M., {Kirby}, E.~N., \& {Simon}, J.~D. 2013, \apj, 767,
  134

\bibitem[{{Wolf} {et~al.}(2010){Wolf}, {Martinez}, {Bullock}, {Kaplinghat},
  {Geha}, {Mu{\~n}oz}, {Simon}, \& {Avedo}}]{wolf10}
{Wolf}, J., {Martinez}, G.~D., {Bullock}, J.~S., {Kaplinghat}, M., {Geha}, M.,
  {Mu{\~n}oz}, R.~R., {Simon}, J.~D., \& {Avedo}, F.~F. 2010, \mnras, 406, 1220

\bibitem[{{Woosley} \& {Weaver}(1995)}]{woosley95}
{Woosley}, S.~E., \& {Weaver}, T.~A. 1995, \apjs, 101, 181

\bibitem[{{York} {et~al.}(2000){York}, {Adelman}, {Anderson}, {Anderson},
  {Annis}, {Bahcall}, {Bakken}, {Barkhouser}, {Bastian}, {Berman}, {Boroski},
  {Bracker}, {Briegel}, {Briggs}, {Brinkmann}, {Brunner}, {Burles}, {Carey},
  {Carr}, {Castander}, {Chen}, {Colestock}, {Connolly}, {Crocker}, {Csabai},
  {Czarapata}, {Davis}, {Doi}, {Dombeck}, {Eisenstein}, {Ellman}, {Elms},
  {Evans}, {Fan}, {Federwitz}, {Fiscelli}, {Friedman}, {Frieman}, {Fukugita},
  {Gillespie}, {Gunn}, {Gurbani}, {de Haas}, {Haldeman}, {Harris}, {Hayes},
  {Heckman}, {Hennessy}, {Hindsley}, {Holm}, {Holmgren}, {Huang}, {Hull},
  {Husby}, {Ichikawa}, {Ichikawa}, {Ivezi{\'c}}, {Kent}, {Kim}, {Kinney},
  {Klaene}, {Kleinman}, {Kleinman}, {Knapp}, {Korienek}, {Kron}, {Kunszt},
  {Lamb}, {Lee}, {Leger}, {Limmongkol}, {Lindenmeyer}, {Long}, {Loomis},
  {Loveday}, {Lucinio}, {Lupton}, {MacKinnon}, {Mannery}, {Mantsch}, {Margon},
  {McGehee}, {McKay}, {Meiksin}, {Merelli}, {Monet}, {Munn}, {Narayanan},
  {Nash}, {Neilsen}, {Neswold}, {Newberg}, {Nichol}, {Nicinski}, {Nonino},
  {Okada}, {Okamura}, {Ostriker}, {Owen}, {Pauls}, {Peoples}, {Peterson},
  {Petravick}, {Pier}, {Pope}, {Pordes}, {Prosapio}, {Rechenmacher}, {Quinn},
  {Richards}, {Richmond}, {Rivetta}, {Rockosi}, {Ruthmansdorfer}, {Sandford},
  {Schlegel}, {Schneider}, {Sekiguchi}, {Sergey}, {Shimasaku}, {Siegmund},
  {Smee}, {Smith}, {Snedden}, {Stone}, {Stoughton}, {Strauss}, {Stubbs},
  {SubbaRao}, {Szalay}, {Szapudi}, {Szokoly}, {Thakar}, {Tremonti}, {Tucker},
  {Uomoto}, {Vanden Berk}, {Vogeley}, {Waddell}, {Wang}, {Watanabe},
  {Weinberg}, {Yanny}, {Yasuda}, \& {SDSS Collaboration}}]{york00}
{York}, D.~G., {et~al.} 2000, \aj, 120, 1579

\bibitem[{{Yoshida} {et~al.}(2008){Yoshida}, {Omukai}, \&
  {Hernquist}}]{yoshida08}
{Yoshida}, N., {Omukai}, K., \& {Hernquist}, L. 2008, Science, 321, 669

\end{thebibliography}


\newpage






\end{document}